%% file: HI_M33.tex
\documentclass[twocolumn]{aastex6}

\pdfoutput=1 

\usepackage{footnote}
\usepackage{graphicx}
\usepackage{longtable}
\usepackage{txfonts}

\newcommand{\hi}{\mbox{H{\sc i}}}
\newcommand{\hI}{\mbox{H{\sc i}}}
\newcommand{\hii}{\mbox{H{\textsc{ii}}}}
\newcommand{\hII}{\mbox{H{\textsc{ii}}}}
\newcommand{\ha}{H$\alpha$}
\newcommand {\hdeux} {H$\mathrm{_2}$}

\newcommand{\kms}{km s$^{-1}$}

\newcommand{\msol}{\rm M$_\odot$}

\newcommand{\lsol}{$L_\odot$}

\slugcomment{Accepted for publication in the  Astronomical Journal}

\shorttitle{\hi\ Kinematics and Mass Distribution of Messier 33}
\shortauthors{Kam et al.}

\begin{document}
 
 \title {\hi\ Kinematics and Mass Distribution of Messier 33}
  
 \author{S. Z. Kam \altaffilmark{1,2}, C. Carignan\altaffilmark{1,2,3}, L. Chemin\altaffilmark{4,5},  T. Foster\altaffilmark{6},
 E. Elson\altaffilmark{3}, T. H. Jarrett\altaffilmark{3}}
  
 \altaffiltext{1}{Laboratoire de Physique et de Chimie de l'Environnement,  Observatoire d'Astrophysique de l'Universit\'e  Ouaga I Pr Joseph Ki-Zerbo (ODAUO), 
 03 BP  7021, Ouaga 03, Burkina Faso; \textbf{email: szachkam@gmail.com}}

\altaffiltext{2}{D\'epartement de physique, Universit\'e de Montr\'eal, C.P. 6128, Succ. centre-ville, Montr\'eal, Qu\'ebec, Canada, H3C 3J7}
\altaffiltext{3}{Department of Astronomy, University of Cape Town, Private Bag X3, Rondebosch 7701, South Africa; \textbf{email: ccarignan@ast.uct.ac.za}}
  
\altaffiltext{4}{INPE/MCT, Divis\~ao de Astrof\'isica, Av. dos Astronautas, 1.758, S\~ao Jose dos Campos, SP, 12227-010, Brazil}
\altaffiltext{5}{Universidad de Antofagasta, Unidad de Astronom\'ia, Antofagasta, Avenida Angamos 601, Antofagasta 1270300, Chile; \textbf{email: laurent.chemin@uantof.cl}}

\altaffiltext{6}{Dominion Radio Astrophysical Observatory,  P.O. Box 248, Penticton, British Columbia, V2A 6J9, Canada}

\begin{abstract}
 A new deep  \hi\ survey of  the galaxy Messier 33 is 
  presented, based on observations obtained at the Dominion Radio Astrophysical 
 Observatory.   We observe a perturbed outer gas distribution and kinematics in M33, and confirm the 
  disk warping as a significant  kinematical twist of the major axis of the velocity field, though no strong tilt is measured, 
  in agreement with previous  work.
 Evidence for a new low brightness \hi\  component with anomalous velocity is reported.  
  It harbours a large velocity scatter, as its kinematics both exceeds and lags the rotation of the disk, 
  and leaks in the forbidden velocity zone of apparent counter-rotation.
   The   observations also reveal  wide and multiple peak \hi\ profiles  
   which can be partly explained by crowded  orbits in the framework of the warp model. 
 Asymmetric motions are identified in the velocity field, 
  as possible signatures of a lopsided potential and  the warp. 
   The mass distribution modeling of the hybrid \ha-\hi\ rotation curve favours a cuspy dark matter halo with a
    concentration in disagreement with the $\Lambda$CDM dark halo mass-concentration relationship. 
  The total mass  enclosed in 23 kpc is $8\, 10^{10}$  \msol,  
   of which  11\% are stars and gas. At the  virial radius of the cuspy halo, 
  the resulting total mass is  $5\, 10^{11}$ \msol, but with a baryonic 
  mass fraction of 2\% only. This strongly suggests a more realistic   radius  encompassing the total mass  of M33 
  well smaller than  the virial radius  of the halo, maybe comparable to the size of the \hi\ disk.  
 \end{abstract}
 

\maketitle

\section{Introduction}
	\label{sec:intro} 
	\input{Introductionhi} 
\section{21-cm observations and data reduction}
	\label{sec:obs}
	\input{Observation}

\section{\hi\ distribution and kinematics}
	\label{sec:velo}
	\input{Velocity}	
\section{Mass distribution models of Messier 33}
	\label{sec:massmod}
	\input{Massmodel}
\section{Conclusion}
	\label{sec:conclusion}
	\input{Conclusion}

\section*{Acknowledgements}
The Dominion Radio Astrophysical Observatory is operated as a national facility by the National Research Council of Canada. ZKS's work was supported by CC's Discovery grant of the Natural Sciences and Engineering Research Council of Canada.
The work of CC and THJ is based upon research supported by the South African Research Chairs Initiative (SARChI) of the Department of Science and Technology (DST),  the Square Kilometer Array South Africa (SKA SA) and the National Research Foundation (NRF).
The research of EE has been supported by an SKA SA fellowship.
LC acknowledges financial supports from CNPq (PCI/INPE) project 301176/2016-7, Comit\'e Mixto ESO-Chile, and 
DGI University of Antofagasta.  
TF's work has partially been supported by a grant from the Brandon University  Research Committee (BURC).
We  are very grateful to Kevin Douglas 
and Mary Putman for providing us the Arecibo TOGS data, to 
Cl\'ement Druard for having provided the emission 
map of the molecular gas, to Pierre Gratier 
for sharing the high-resolution VLA datacube, and to 
Jonathan Braine for his comments that greatly helped improved 
the analysis and the manuscript.

\appendix

\section{\hi\ dispersion and mass surface density of M33}
\label{sec:siglossurfdens}
 
\begin{table*}
\begin{center}
 \label{tab:siglossurfdens}
\caption{\hi\ dispersion and mass surface density of M33}
\begin{tabular}{ccc|ccc|ccc|ccc}
\hline\hline 
 Radius &   $\rm \sigma_{los} $ & $\rm \Sigma_{H\textsc{i}}$  &  Radius     & $\rm \sigma_{los}$ &  $\rm \Sigma_{H\textsc{i}}$  & 
 Radius     & $\rm \sigma_{los}$ &  $\rm \Sigma_{H\textsc{i}}$ &  Radius     & $\rm \sigma_{los}$ &  $\rm \Sigma_{H\textsc{i}}$  \\
\hline
     2  &  0.7 & 7.89 &  26  &  7.6 & 7.54 &   50  & 8.5 & 0.49 &  74   & 8.2 & 0.03  \\
     4  &  9.5 & 7.73 &  28  &  7.6 & 6.50 &   52  & 8.2 & 0.36 &  76   & 6.5 & 0.04  \\
     6  &  9.3 & 8.26 &  30  &  8.0 & 4.99 &   54  & 8.3 & 0.23 &  78   & 6.5 & 0.04  \\
     8  &  9.6 & 8.52 &  32  &  8.9 & 3.50 &   56  & 7.6 & 0.18 &  80   & 7.4 & 0.04  \\
    10  &  0.0 & 8.23 &  34  &  9.8 & 2.51 &   58  & 7.7 & 0.12 &  82   & 6.5 & 0.08  \\
    12  &  0.1 & 7.66 &  36  &  0.2 & 1.75 &   60  & 6.8 & 0.08 &  84   & 5.6 & 0.09  \\
    14  &  9.1 & 8.00 &  38  &  0.3 & 1.34 &   62  & 7.4 & 0.04 &  86   & 6.1 &	\nodata  \\
    16  &  8.2 & 8.61 &  40  &  0.5 & 1.00 &   64  & 7.5 & 0.03 &  88   & 5.7 & \nodata  \\
    18  &  8.0 & 8.14 &  42  &  9.5 & 0.75 &   66  & 6.7 & 0.03 &  90   & 5.7 & 0.09  \\
    20  &  8.4 & 8.19 &  44  &  8.8 & 0.67 &   68  & 6.6 & 0.00 &  92   & 5.6 & \nodata  \\
    22  &  7.9 & 7.61 &  46  &  9.1 & 0.56 &   70  & 6.4 & 0.03 &  94   & 6.4 & \nodata  \\
    24  &  8.0 & 7.57 &  48  &  9.0 & 0.63 &   72  & 8.0 & 0.07 &  96   & 7.5 & \nodata  \\
\hline 
\end{tabular} 
\end{center}
Comments: The radius is in arcmin, the  line-of-sight  dispersion $\rm \sigma_{los} $ in \kms, and the surface density 
$\rm \Sigma_{H\textsc{i}}$ in \msol\, pc$^{-2}$.
\end{table*}

\section{Hybrid \ha-\hi\ rotation curve of M33}
\label{sec:hybridrc}

\begin{deluxetable*}{ccc|ccc|ccc|ccc|ccc}
\tablecaption{Hybrid \ha-\hi\ rotation curve of M33}
\tablehead{
 $\ln(R)$  & $\rm V_{rot}$ & $\rm \Delta V_{rot}$ & $\ln(R)$  & $\rm V_{rot}$ & $\rm \Delta V_{rot}$ & 
 $\ln(R)$  & $\rm V_{rot}$ & $\rm \Delta V_{rot}$ & $\ln(R)$  & $\rm V_{rot}$ & $\rm \Delta V_{rot}$ & 
 $\ln(R)$  & $\rm V_{rot}$ & $\rm \Delta V_{rot}$ }
\startdata
1.61 &   6.20 &  2.32 & 5.87 &  71.95 &  4.15 & 6.56 &  85.63 &  0.99 & 6.96 &  97.13 &  0.56 & 7.25 & 108.25 &  6.96 \\
2.30 &   9.30 &  2.71 & 5.89 &  72.19 &  4.56 & 6.57 &  84.64 &  0.62 & 6.97 & 100.87 &  1.88 & 7.25 & 107.50 &  7.89 \\
2.71 &  13.00 &  0.55 & 5.90 &  72.40 &  5.20 & 6.57 &  87.03 &  0.66 & 6.97 & 100.32 &  2.41 & 7.25 & 108.45 &  5.87 \\
3.00 &  18.25 &  3.79 & 5.91 &  71.80 &  5.68 & 6.58 &  89.26 &  1.30 & 6.98 &  97.17 &  0.57 & 7.26 & 108.88 &  5.48 \\
3.22 &  20.73 &  3.31 & 5.93 &  71.14 &  4.69 & 6.59 &  90.95 &  1.33 & 6.98 &  98.41 &  1.45 & 7.26 & 107.92 &  5.56 \\
3.40 &  22.69 &  3.50 & 5.94 &  71.81 &  4.74 & 6.59 &  92.36 &  1.26 & 6.98 &  98.67 &  0.55 & 7.27 & 105.22 &  6.43 \\
3.56 &  24.54 &  1.37 & 5.95 &  71.89 &  7.37 & 6.60 &  92.75 &  0.66 & 6.99 & 101.05 &  0.62 & 7.27 & 105.77 &  6.42 \\
3.69 &  26.23 &  0.68 & 5.97 &  72.47 &  7.82 & 6.61 &  91.76 &  2.36 & 6.99 & 101.51 &  0.55 & 7.27 & 110.49 &  4.32 \\
3.81 &  30.62 &  1.15 & 5.98 &  73.25 &  7.98 & 6.61 &  91.09 &  3.59 & 7.00 & 102.93 &  1.20 & 7.28 & 110.51 &  4.05 \\
3.91 &  31.84 &  0.94 & 5.99 &  73.46 &  8.12 & 6.62 &  91.89 &  2.46 & 7.00 & 103.97 &  1.17 & 7.28 & 105.47 &  5.25 \\
4.01 &  34.51 &  1.18 & 6.00 &  74.46 &  8.06 & 6.63 &  91.47 &  2.86 & 7.01 & 103.50 &  1.60 & 7.28 & 105.33 &  5.86 \\
4.09 &  35.42 &  1.46 & 6.02 &  75.18 &  5.68 & 6.63 &  92.49 &  1.66 & 7.01 & 104.44 &  1.15 & 7.29 & 109.38 &  1.96 \\
4.17 &  37.37 &  2.00 & 6.03 &  76.20 &  4.64 & 6.64 &  92.69 &  1.02 & 7.02 & 104.81 &  1.10 & 7.29 & 113.28 &  4.31 \\
4.25 &  38.38 &  2.49 & 6.04 &  77.74 &  3.39 & 6.65 &  93.88 &  0.66 & 7.02 & 104.98 &  1.67 & 7.29 & 115.69 &  2.75 \\
4.32 &  39.20 &  1.20 & 6.05 &  78.30 &  2.56 & 6.65 &  95.00 &  0.86 & 7.03 & 105.49 &  1.90 & 7.30 & 116.42 &  2.97 \\
4.38 &  41.82 &  2.47 & 6.06 &  78.07 &  2.50 & 6.66 &  94.41 &  0.56 & 7.03 & 106.07 &  1.72 & 7.30 & 115.31 &  3.52 \\
4.44 &  41.08 &  2.43 & 6.08 &  77.82 &  3.33 & 6.67 &  93.85 &  2.13 & 7.03 & 106.00 &  0.98 & 7.30 & 112.56 &  3.18 \\
4.50 &  41.74 &  1.71 & 6.09 &  78.52 &  2.93 & 6.67 &  93.11 &  3.81 & 7.04 & 107.46 &  0.76 & 7.31 & 111.32 &  2.71 \\
4.55 &  42.57 &  2.18 & 6.10 &  78.85 &  2.04 & 6.68 &  94.65 &  2.99 & 7.04 & 109.03 &  1.80 & 7.31 & 112.61 &  2.99 \\
4.61 &  43.40 &  4.11 & 6.11 &  78.78 &  1.12 & 6.68 &  96.50 &  2.17 & 7.05 & 109.25 &  2.05 & 7.31 & 113.70 &  2.34 \\
4.65 &  43.85 &  3.64 & 6.12 &  78.14 &  0.57 & 6.69 &  96.79 &  1.76 & 7.05 & 108.48 &  0.71 & 7.32 & 114.91 &  0.89 \\
4.70 &  45.95 &  3.03 & 6.13 &  79.82 &  1.28 & 6.70 &  96.32 &  0.60 & 7.06 & 107.45 &  1.20 & 7.32 & 115.79 &  5.15 \\
4.74 &  48.40 &  0.96 & 6.14 &  79.36 &  1.60 & 6.70 &  96.00 &  1.00 & 7.06 & 107.63 &  1.50 & 7.32 & 112.92 &  9.94 \\
4.79 &  49.04 &  2.33 & 6.15 &  79.58 &  1.10 & 6.71 &  95.99 &  0.55 & 7.06 & 107.73 &  1.95 & 7.33 & 119.07 &  5.09 \\
4.83 &  49.37 &  4.79 & 6.16 &  79.11 &  0.55 & 6.72 &  94.58 &  0.82 & 7.07 & 108.36 &  0.88 & 7.33 & 111.82 & 13.33 \\
4.87 &  49.58 &  5.19 & 6.17 &  79.62 &  0.88 & 6.72 &  95.24 &  0.56 & 7.07 & 108.79 &  0.61 & 7.33 & 113.09 & 13.94 \\
4.91 &  50.83 &  3.94 & 6.18 &  82.16 &  2.31 & 6.73 &  95.08 &  0.57 & 7.08 & 108.59 &  0.70 & 7.34 & 110.64 & 13.95 \\
4.94 &  52.27 &  4.79 & 6.19 &  82.16 &  2.24 & 6.73 &  96.30 &  2.29 & 7.08 & 108.83 &  0.71 & 7.34 & 108.71 & 15.32 \\
4.98 &  52.64 &  6.30 & 6.20 &  82.59 &  3.54 & 6.74 &  96.56 &  1.87 & 7.09 & 104.87 &  3.58 & 7.34 & 106.90 & 18.46 \\
5.01 &  52.61 &  7.27 & 6.21 &  82.91 &  4.06 & 6.75 &  96.82 &  0.55 & 7.09 & 103.69 &  3.97 & 7.35 & 106.47 & 17.74 \\
5.04 &  52.97 &  6.74 & 6.22 &  82.02 &  3.83 & 6.75 &  98.17 &  0.70 & 7.09 & 106.19 &  1.31 & 7.35 & 107.07 & 17.69 \\
5.08 &  54.45 &  5.94 & 6.23 &  81.00 &  4.64 & 6.76 &  97.69 &  1.36 & 7.10 & 107.36 &  0.62 & 7.35 & 107.73 & 19.17 \\
5.11 &  54.75 &  6.23 & 6.24 &  81.58 &  3.35 & 6.76 &  96.95 &  0.65 & 7.10 & 107.99 &  1.15 & 7.36 & 110.16 & 17.09 \\
5.14 &  53.44 &  5.79 & 6.25 &  82.93 &  4.11 & 6.77 &  95.99 &  0.64 & 7.11 & 108.06 &  0.70 & 7.36 & 109.82 & 17.98 \\
5.16 &  55.63 &  4.15 & 6.26 &  83.81 &  4.91 & 6.77 &  95.19 &  0.90 & 7.11 & 107.26 &  1.15 & 7.36 & 111.36 & 19.37 \\
5.19 &  55.99 &  1.39 & 6.27 &  83.69 &  5.27 & 6.78 &  94.76 &  2.86 & 7.11 & 107.30 &  1.04 & 7.37 & 108.01 & 16.48 \\
5.22 &  54.21 &  2.04 & 6.28 &  82.93 &  4.98 & 6.79 &  94.75 &  1.55 & 7.12 & 105.96 &  1.11 & 7.37 & 104.27 & 15.46 \\
5.25 &  52.71 &  3.09 & 6.29 &  81.05 &  2.57 & 6.79 &  94.29 &  2.72 & 7.12 & 106.36 &  0.91 & 7.37 & 104.57 & 16.63 \\
5.27 &  52.96 &  2.03 & 6.30 &  80.03 &  0.96 & 6.80 &  95.88 &  4.38 & 7.13 & 107.76 &  1.47 & 7.37 & 107.86 & 17.65 \\
5.30 &  55.06 &  0.83 & 6.31 &  79.84 &  1.28 & 6.80 &  98.05 &  6.72 & 7.13 & 106.64 &  0.70 & 7.43 & 106.76 &  2.16 \\
5.32 &  55.61 &  1.61 & 6.32 &  78.11 &  0.64 & 6.81 &  98.67 &  6.77 & 7.13 & 105.83 &  1.00 & 7.50 & 107.33 &  3.02 \\
5.35 &  56.01 &  1.60 & 6.33 &  78.30 &  0.80 & 6.81 &  99.73 &  8.12 & 7.14 & 107.77 &  1.97 & 7.56 & 108.29 &  4.01 \\
5.37 &  56.37 &  1.47 & 6.34 &  79.95 &  1.11 & 6.82 & 100.12 &  8.06 & 7.14 & 109.46 &  0.58 & 7.62 & 109.72 &  4.01 \\
5.39 &  57.87 &  0.86 & 6.35 &  81.21 &  2.43 & 6.82 & 101.28 &  8.81 & 7.15 & 111.04 &  0.96 & 7.68 & 111.98 &  4.78 \\
5.42 &  58.99 &  0.80 & 6.35 &  82.36 &  4.01 & 6.83 & 100.50 &  8.65 & 7.15 & 105.97 &  2.53 & 7.73 & 116.06 &  2.18 \\
5.44 &  59.17 &  1.33 & 6.36 &  81.44 &  3.53 & 6.84 & 100.15 &  8.55 & 7.15 & 104.96 &  5.49 & 7.78 & 117.23 &  2.45 \\
5.46 &  60.13 &  2.08 & 6.37 &  80.95 &  4.15 & 6.84 & 102.06 &  8.80 & 7.16 & 111.44 &  8.30 & 7.83 & 116.46 &  6.48 \\
5.48 &  60.68 &  1.47 & 6.38 &  81.85 &  3.98 & 6.85 &  98.96 &  5.91 & 7.16 & 112.71 &  7.21 & 7.88 & 115.68 &  8.07 \\
5.50 &  61.23 &  2.09 & 6.39 &  81.83 &  2.81 & 6.85 &  96.91 &  3.30 & 7.17 & 112.66 &  7.25 & 7.92 & 117.40 &  8.23 \\
5.52 &  61.59 &  1.58 & 6.40 &  81.39 &  3.26 & 6.86 &  97.79 &  4.86 & 7.17 &  99.94 &  4.97 & 7.97 & 116.84 &  8.93 \\
5.54 &  61.26 &  1.64 & 6.41 &  81.60 &  3.75 & 6.86 &  97.54 &  4.49 & 7.17 &  95.60 &  9.12 & 8.01 & 115.70 &  9.64 \\
5.56 &  62.38 &  2.57 & 6.41 &  83.01 &  1.77 & 6.87 &  96.80 &  3.84 & 7.18 &  97.16 &  9.68 & 8.05 & 115.09 &  7.69 \\
5.58 &  63.79 &  3.62 & 6.42 &  82.73 &  1.34 & 6.87 &  96.34 &  3.32 & 7.18 &  97.93 &  4.86 & 8.08 & 117.09 &  5.11 \\
5.60 &  64.71 &  4.29 & 6.43 &  82.19 &  2.18 & 6.88 &  96.61 &  2.97 & 7.19 & 102.58 &  5.33 & 8.12 & 118.20 &  3.15 \\
5.62 &  64.44 &  4.18 & 6.44 &  83.79 &  0.84 & 6.88 &  96.58 &  2.36 & 7.19 & 103.68 &  4.01 & 8.15 & 118.42 &  1.42 \\
5.63 &  64.76 &  5.09 & 6.45 &  84.23 &  2.02 & 6.89 &  97.27 &  3.13 & 7.19 & 104.75 &  4.99 & 8.19 & 118.20 &  1.78 \\
5.65 &  66.67 &  4.51 & 6.45 &  85.03 &  4.17 & 6.89 &  96.27 &  3.97 & 7.20 & 106.04 &  7.02 & 8.22 & 117.45 &  2.38 \\
5.67 &  68.62 &  3.97 & 6.46 &  86.46 &  2.47 & 6.90 &  97.57 &  4.77 & 7.20 & 106.78 &  5.93 & 8.25 & 119.56 &  0.79 \\
5.69 &  68.09 &  3.94 & 6.47 &  86.55 &  1.43 & 6.90 &  99.50 &  4.53 & 7.20 & 105.56 &  6.01 & 8.28 & 118.60 &  1.53 \\
5.70 &  65.49 &  4.83 & 6.48 &  86.27 &  0.82 & 6.91 & 103.83 &  7.82 & 7.21 & 104.72 &  4.71 & 8.31 & 122.63 &  0.50 \\
5.72 &  65.61 &  5.44 & 6.48 &  86.51 &  0.77 & 6.91 & 103.79 &  7.01 & 7.21 & 106.75 &  4.59 & 8.34 & 124.10 &  2.86 \\
5.74 &  66.85 &  4.35 & 6.49 &  86.51 &  1.19 & 6.92 & 103.69 &  5.59 & 7.22 & 107.06 &  5.78 & 8.37 & 125.03 &  2.19 \\
5.75 &  67.49 &  5.55 & 6.50 &  86.73 &  2.25 & 6.92 & 102.16 &  4.12 & 7.22 & 104.89 &  6.04 & 8.40 & 125.49 &  2.54 \\
5.77 &  67.81 &  5.14 & 6.51 &  85.88 &  1.65 & 6.93 &  99.43 &  3.12 & 7.22 & 106.56 &  4.29 & 8.42 & 125.24 &  8.08 \\
5.78 &  68.85 &  4.03 & 6.51 &  85.95 &  0.60 & 6.93 &  99.70 &  2.12 & 7.23 & 107.95 &  4.16 & 8.45 & 121.96 &  9.76 \\
5.80 &  68.92 &  4.30 & 6.52 &  85.53 &  0.55 & 6.94 & 103.82 &  6.87 & 7.23 & 109.25 &  3.79 & 8.48 & 120.39 &  8.47 \\
5.81 &  70.75 &  3.98 & 6.53 &  85.95 &  2.42 & 6.94 &  98.75 &  0.60 & 7.23 & 110.55 &  3.30 & 8.50 & 114.04 & 26.64 \\
5.83 &  71.41 &  3.79 & 6.54 &  85.40 &  1.30 & 6.95 &  99.30 &  1.32 & 7.24 & 111.55 &  3.30 & 8.52 & 110.03 & 34.64 \\
5.84 &  71.66 &  4.62 & 6.54 &  82.53 &  0.55 & 6.95 &  99.55 &  1.55 & 7.24 & 110.80 &  1.73 & 8.55 &  98.69 & 27.45 \\
5.86 &  72.38 &  4.01 & 6.55 &  85.17 &  0.67 & 6.96 & 104.52 &  5.50 & 7.24 & 110.85 &  2.90 & 8.57 & 100.07 & 33.42 \\
8.59 & 104.32 & 35.17 & 8.62 & 101.19 & 27.35 & 8.64 & 123.49 & 39.13 &      &        &       &      &        &       \\
\enddata
~\\ Comments: $\ln(R)$ is the neperian logarithm of the radius (radius in arcsec unit). The rotation velocity and its  uncertainty are in \kms.
 \label{tab:hybridrc}
\end{deluxetable*}

\end{document}

%% file: Introductionhi.tex
With the Milky Way and  Andromeda, the Triangulum galaxy (Messier 33, the third most massive disk galaxy of the Local Group) 
has long been among the most studied nearby galaxies to scrutinize the chemical, dynamical 
and structural properties of the stellar populations and of
the interstellar medium.
  In particular, as it is a prototype of gas-rich  spirals of moderate inclination (Table~\ref{tab:oparam}), 
it is very appropriate to study the relationships of  the atomic, molecular and ionized gas content with  star formation inside the disk.  
Since it is now well admitted that M33 has undergone a tidal encounter with his massive companion M31,
 as shown by the perturbation of its close environment 
\citep[e.g.][]{Braun2004,McConnachie2010,Wolfe2013}, it is expected that gas expelled during that
interaction is currently returning into the M33  disk, fueling the active star formation \citep{Putman2009}. 

The implied important population of \hII\ regions 
\citep[e.g.][]{Boulesteix1974,Zaritsky1989, Relano2013} 
has thus been a motivation for us to present the first large-scale,
arcsec-resolution, 3D spectroscopy survey of the disk of M33 in the \ha\ emission line \citep[][hereafter Kam15]{ZKam2015}.
On one hand, the objectives of this survey were to measure the internal
kinematics of star forming regions.  
For instance, Kam15 presented detailed velocity fields of compact and extremely large \hii\ regions, like NGC 604, 
or the relation between the velocity dispersion and the integrated intensity of the \ha\ line, 
underlying the physical processes occurring in these regions and 
in the diffuse interstellar medium (stellar winds, expansion, etc.). The catalog of \hii\ regions to be provided from this survey 
will also be useful to study the relationships with  other tracers of star formation in M33 at an unprecedented level of details.
On the other hand,  the \ha\ mapping has been a unique opportunity 
to determine for the first time  the most extended \ha\ velocity field of M33. 
 The modeling of the velocity field led Kam15  to conclude that 
 the kinematical parameters of the ionized gas disk are very consistent 
 with those of the stellar disk.  
 The most important result of Kam15 is the determination of 
 the \ha\ rotation curve out to 8 kpc sampled every 20 pc, which is, as of date, the 
 most resolved rotation curve obtained for any massive spiral galaxies other than the Milky Way. 
 That curve perfectly traces the velocity gradient in the inner disk,
 and presents many wiggles characteristic of spiral arms perturbations, 
 barely seen in previous \hi\ and CO observations of M33. 
 These irregularities however do not prevent the \ha, \hi\ or CO rotation velocities to remain in good agreement.  
 
 The derivation of the \ha\ rotation curve  actually constitutes the first pillar of a broader project
 devoted to revisit the modeling of the mass distribution of M33. 
 The other pillar inherent to such modeling is to get the \hi\ rotation curve to cover as far as possible the outer disk, 
 in order  to obtain accurate fits of the dark matter (DM) distribution, 
 or of alternate gravity models such as Modified Newtonian Dynamics 
 \citep[MOND,][]{Milgrom1983b,Milgrom1983a}.  
 
 Existing \hi\ studies of M33 attest to which extent the tidal interaction with Andromeda has not been without consequences on the gas distribution in the disk outskirts, 
 revealing perturbed  features like the strong warp,
 arc-like structures or diffuse and discrete gas around the disk  \citep[][]{Corbelli1997,Putman2009,Lockman2012}.  
 Recently, \cite{Corbelli2014} combined VLA and GBT observations to model the \hi\ warp and rotation curve.  
 They have presented a new rotation curve that extends about 5 kpc further out than previous studies. 
 They have also shown that the distribution of dark matter   is consistent with a model for which
 the mass density steeply decreases in the centre of the halo \citep[the cosmological cusp \`a la Navarro-Frenk-White, see][]{Navarro1997}, 
and whose concentration agrees well with the halo mass-concentration relation from $\Lambda$ Cold Dark Matter simulations \citep{Ludlow2014}. 
 Comparable results have been obtained by \cite{hague2015}, but from a lower resolution \hi\  curve derived earlier by \cite{Corbelli2000}. 
They concluded  that models with density profiles with an inner slope shallower than  0.9 (as measured at $R \sim 0.5$ kpc, the first velocity point of their rotation curve) 
 are only compatible with 3.6-$\rm \mu m$ mass-to-light ratios $\Upsilon > 2$, while cuspier densities can coexist with $\Upsilon <2$. 
 However, a halo with a constant density core in M33 would be difficult to reconcile with stellar populations models, or with other observational 
 large scale dynamical studies \citep[e.g.][]{Lelli2016}. This is something we want to revisit with a new set of data.
 
 In this context, we have performed an \hi\ survey of M33 at the Dominion Radio Astrophysical Observatory (DRAO). 
This arcminute-resolution survey   is a good intermediate between   VLA and Arecibo or GBT measurements. 
 The objectives of this article are first to  present the survey, the gas content and distribution of M33, a new tilted-ring model of the \hi\ velocity field and 
 the \hi\ rotation curve.  We also want to  examine the shape and amplitude of the outer rotation curve, at radii beyond $R=17$ kpc 
 where \citet{Corbelli2014} presented new velocities. 
  The second objective of the article is to benefit from the high-resolution  survey of Kam15 and 
 perform   mass distribution models from an hybrid resolution \ha-\hi\ rotation curve. 
 In particular we want to determine the most appropriate stellar mass, density profile  of dark matter, 
 and infer the total mass of M33. In addition, we want to compare dark matter mass models with Modified Newtonian Dynamics.

 Throughout the article, we adopt a Hubble constant of 68 \kms\ Mpc$^{-1}$ \citep{planck16} and a distance to M33 of 0.84 Mpc (see Kam15 and references therein).
 The basic parameters of M33 are summarized in Table~\ref{tab:oparam}. 
 The  observations and reduction of the new DRAO data are presented in Section~\ref{sec:obs}, which also  
 gives general characteristics of a combined DRAO+Arecibo \hi\ datacube.
  Section~\ref{sec:velo} presents the analysis of the \hi\ distribution, 
 the tilted-ring and Fourier models  of the \hi\ velocity field. It also determines 
 a hybrid \ha-\hi\ rotation curve to be used for the 
 modeling of the mass distribution performed in Section~\ref{sec:massmod}. 
 \begin{table}
\centering
\caption{Parameters of M 33.}
\begin{tabular}{@{}llllllllllllll@{}}
\hline
\hline
Parameters & Value& Source \\ 
\hline
Morphological type	& SA(s)cd & RC3	 \\ 
R.A. (2000) 				& 01$\rm^h$ 33$\rm^m$ 33.1$\rm^s$& RC3 \\
Dec. (2000)				& +30\degr\ 39$^{'}$ 18${''}$& RC3 \\
Systemic Velocity (\kms)		&$-179 \pm3$ & RC3 \\
Distance (Mpc)				& 0.84  &  \\
Scale (pc/arcmin)			& 244 & \\
Disk Scale length (kpc) 		& 1.6 (@ 3.6 $\mu$m) & Kam15 \\
Optical radius, $R_{25}$		& 35$\farcm$4 $\pm$ 1$\farcm$0& RC3 \\
Inclination, $i$				& 52\degr\  $\pm$ 3\degr\ &  WWB \\
Position angle (major axis)	& 22.5\degr\ $\pm$ 1\degr\ &  WWB \\ 	
Apparent magnitude, m$_V$ 	& 5.28& RC3\\ 
Absolute magnitude, M$_V$ 	&$-19.34$& \\
\hline 
 Total \hi\ mass	(\msol)			& $1.95\, 10^9$ & Sec.~\ref{sec:obs} \\
 Systemic Velocity (\kms)			&$-180 \pm3$ 				& Sec.~\ref{sec:obs} \\
 ${\rm V_{rot}}$ maximum (\kms)		& 125 & Sec.~\ref{sec:velo} \\ 
 Stellar mass    (\msol), mass models & $5.5\, 10^9$ &  Sec.~\ref{sec:massmod} \\
 Dynamical mass (\msol), inside $R=23$ kpc 	& $7.9 \, 10^{10}$ &  Sec.~\ref{sec:massmod} \\
\hline
\end{tabular} 

RC3:  \citet{deVaucouleurs1991}; Kam15: \citet{ZKam2015}; WWB: \citet{WWB73}. See Kam15 for the distance to M33, 
as based on a compilation of distance moduli from TRGB, Cepheids and Planetary Nebula Luminosity Function methods.
\label{tab:oparam}
\end{table}

%% file: Observation.tex
\begin{figure}[t]
\includegraphics[width=\columnwidth]{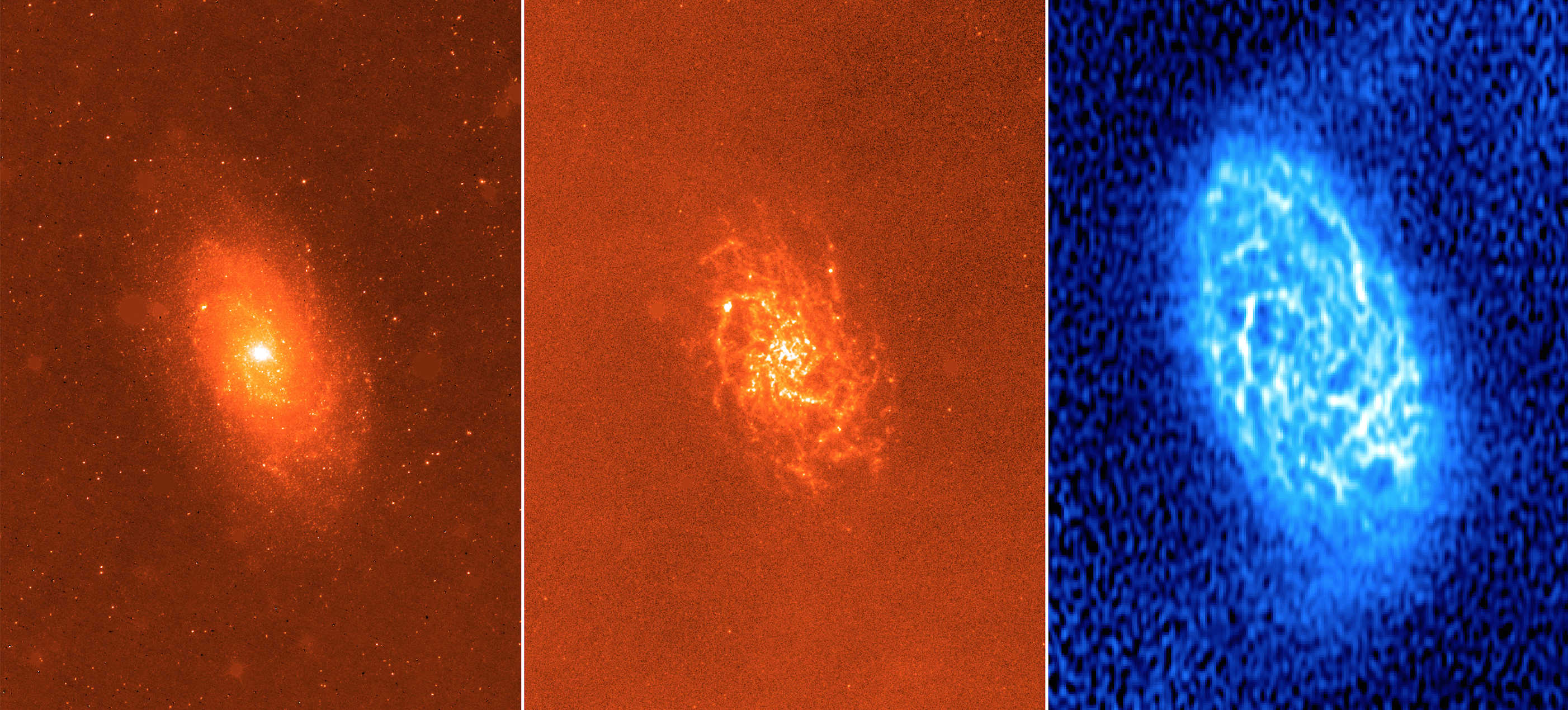}
\caption{WISE W1 (left), W3 (center) and the inner bright \hi\ disk (right) of M33.} 
\label{fig:w1w3hi}
\end{figure}

The primary observations for this study were made with the Synthesis Telescope (ST) at the Dominion Radio Astrophysical Observatory (DRAO). This telescope is an East-West interferometer consisting of seven ($\sim$9~m diameter) dishes spaced variously across a baseline range of 13 to 617~m. At 1420~MHz the 
longest baseline achieves a synthesized half-power beamwidth of 49$\arcsec$(EW) by 49$\arcsec$/sin~$\delta$(NS) with uniform weighting, although in the \hI\  line we use a Gaussian taper in the $u,v$ plane to increase the sensitivity of each velocity channel at the slight expense of resolution (58$\arcsec\times$58$\arcsec/\textrm{sin}\delta$). As the goal of this study is to trace the extended rotation curve of the galaxy, the DRAO instrument was chosen because of two inherent strengths: its wide field (3.1$\degr$), combined with its deep integration (144~h per field). The DRAO instrument is also unmatched in its absolutely calibrated polarization capability \citep[see][for specifications of the DRAO telescope]{Landecker2000,Kothes2010}. Observations of M33 consist of Stokes I,Q,U \& V made in a 30~MHz continuum band centred at 1420~MHz ($\lambda$21~cm), Stokes I in a 2~MHz band at 408~MHz ($\lambda$74~cm), and 256 channels within a 2~MHz band centred on the \hI\  line. For this study we use only the \hI\  data products; future studies will present the total power and polarized radio emission from M33.
The datacube velocity resolution is 2.64 \kms\ at 21-cm and each channel is $\rm \Delta V=1.65$ \kms\
wide. The band was centred on a heliocentric velocity of $\rm V_{\textsc{hel}}=-180$ \kms. Data processing and mosaic-making steps
proceeded as per those in \citet{Chemin2009} for the M31 observations with the DRAO telescope. The measured noise per channel is
$\sim 12-13$ mJy beam$^{-1}$ or $\Delta T_B \sim 1.1$~K in the elliptical synthesized beam (58\arcsec\ $\times$ 114\arcsec).
To increase sensitivity to the faint outermost \hI\  disk of M33 a total of 6 full-synthesis pointings on and around the galaxy were observed and mosaiced together (see Table~\ref{centres}). 

\begin{table}
\centering
\caption{Summary of the six 21~cm \hI\  line synthesis fields centred on and 
surrounding M33, carried out with the DRAO Synthesis Telescope.}
\begin{tabular}{ccc}
\hline
\hline
Observ.&    Field Centre&   Beam Parameters\\
Date&    (RA,~DEC) (J2000.0)& $\rm \theta_{maj} (')\times\theta_{min} (')$, CCWE\\
\hline
09/29/08&
01$^{\textrm{h}}$33$^{\textrm{m}}$50.9$^{\textrm{s}}$,~+30$\degr$39$\arcmin$36$\arcsec$&
$1.90\times 0.97\arcmin$,$-$89.69$\degr$\\
09/29/08&
01$^{\textrm{h}}$36$^{\textrm{m}}$10.2$^{\textrm{s}}$,~+31$\degr$50$\arcmin$34$\arcsec$&
$1.85\times 0.97$,~$-$89.82$\degr$\\
11/05/08&
01$^{\textrm{h}}$31$^{\textrm{m}}$38.2$^{\textrm{s}}$,~+29$\degr$28$\arcmin$12$\arcsec$&
$1.98\times 0.97$,~$-$89.91$\degr$\\
11/05/08&
01$^{\textrm{h}}$34$^{\textrm{m}}$45.8$^{\textrm{s}}$,~+31$\degr$08$\arcmin$13$\arcsec$&
$1.86 \times 0.97$,~$-$90.11$\degr$\\
12/04/08&
01$^{\textrm{h}}$32$^{\textrm{m}}$56.4$^{\textrm{s}}$,~+30$\degr$11$\arcmin$01$\arcsec$&
$1.94 \times 0.97$,~$-$90.30$\degr$\\
12/04/08&
01$^{\textrm{h}}$33$^{\textrm{m}}$50.9$^{\textrm{s}}$,~+30$\degr$39$\arcmin$36$\arcsec$&
1$\arcmin$.91$\times$0$\arcmin$.97,~$-$90.49$\degr$\\
\hline
\end{tabular}

\label{centres}
\end{table}

To recover the large-scale \hI\  structures in M33, we merge single-dish (aka {}``short-spacing'') data obtained from the Turn-On GALFA Survey (TOGS) portion of the GALFA-\hI\  survey at Arecibo. These data, previously 
published by \citet{Putman2009}, have an angular resolution of 3$\farcm$4 and a velocity resolution of 5.15~km~s$^{-1}$, and provide exceptional spatial frequency overlap with the DRAO synthesis data (which are missing structures larger than about 45$\arcmin$). Most importantly, TOGS data are fully corrected for stray-radiation entering the sidelobes, 
thus preventing contamination 
of particularly the M33 outskirts 
 and allowing the extended gas distribution and kinematics to be traced. TOGS \hI\  data are added to the calibrated interferometer-only mosaic in the same manner as in \citet{Chemin2009} after converting them to the same spatial and velocity-resolution and grid. 
 Figure~\ref{fig:w1w3hi} compares the full resolution \hi\ gas distribution to the WISE W1 (3.4 $\mu$m) 
 and W3 (12 $\mu$m) images in which the foreground stars have been identified and removed (Jarrett et al., in prep).
 
 \begin{figure*}[t!]
 \centering
\includegraphics[width=\textwidth]{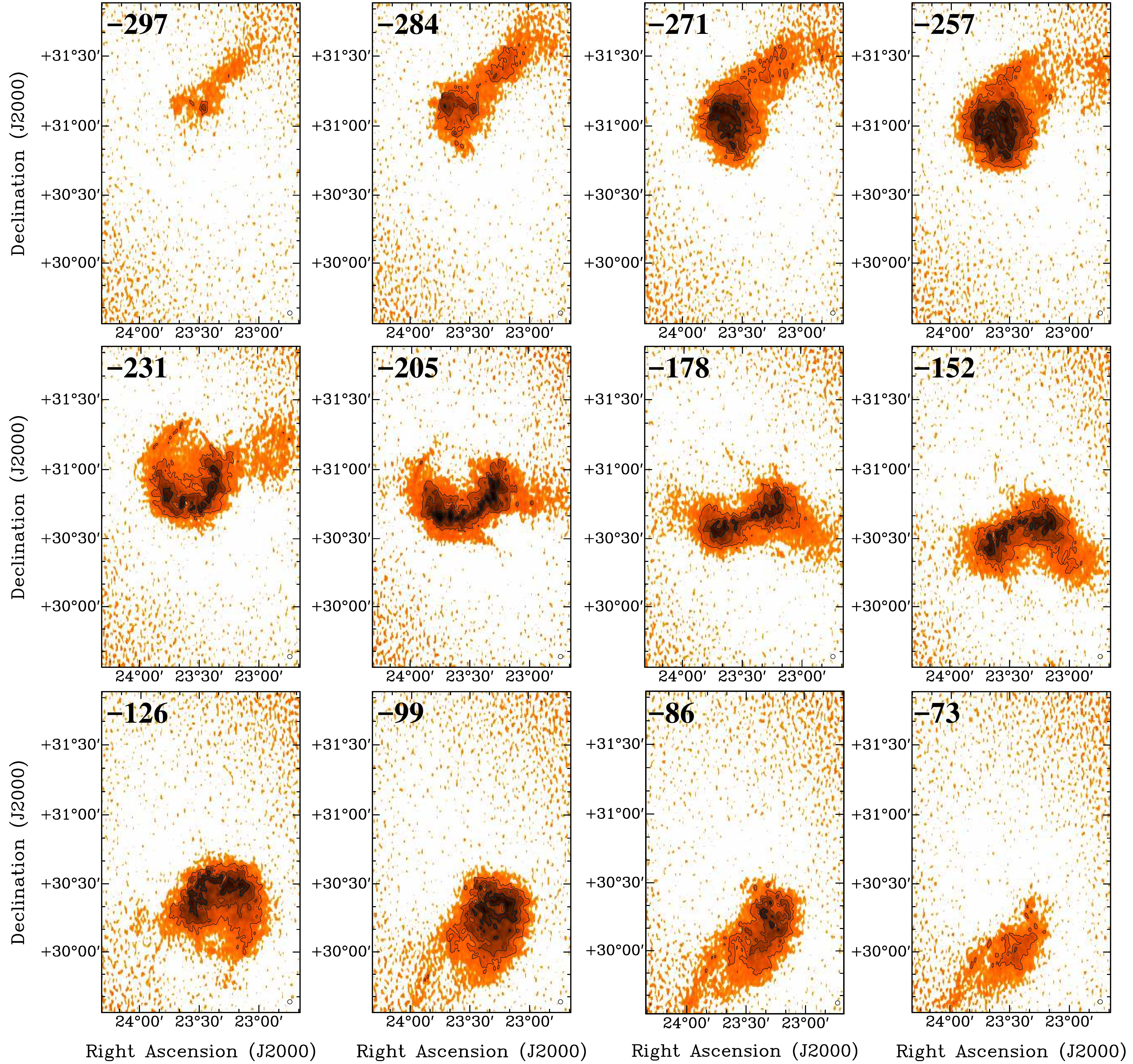}
 \caption{Selected \hi\ channels maps of M33. The brightness temperature scale is logarithmic, with a minimum of $\rm T_B= 0.4$ K.
 Contour levels are for brightness temperatures of 2.5, 6 and 15 K. The circle to the bottom-right corner represents the $2\arcmin \times 2\arcmin$ beam. The heliocentric 
 velocity  of each channel is written on the top-left corner (in \kms). The
 heliocentric velocity of $-178$ \kms\ is the closest to the  
 systemic velocity of M33 ($-180$ \kms) among the selected channels.}
\label{fig:chlmap}
 \end{figure*}
 
To gain sensitivity, we made two final datacubes smoothed to velocity resolutions of 5.3 and 10.6~km~s$^{-1}$, and spatially smoothed
to a circular 120$\arcsec \times$120$\arcsec$ Gaussian beam. 
The measured 1$\sigma$ noise at the pointing centre of the final
short-spacings-added 2$\arcmin$ spatial resolution and 10~km~s$^{-1}$ velocity resolution mosaic is 2~mJy beam$^{-1}$, or $\sim 80$ mK channel$^{-1}$. 

While our final \hI\  cube is as sensitive on a per-beam basis as the VLA BCD data presented in \citet{Gratier2010}
 and in \citet{Corbelli2014}, the 120$\arcsec$ and 10~km~s$^{-1}$ resolution mosaic certainly lacks the
  fine detail and velocity resolution of the current state-of-the-art mapping of M33. 
Rather, it is intended to trace the extended \hI\  disk out to very large radii,
and large \hi\ structures. 
Indeed, since the typical cloud-cloud velocity dispersion in the halo of the Milky Way is 25~km~s$^{-1}$ \citep{deHeij2002}, 
our 10~km~s$^{-1}$ velocity resolution cube is more than sufficient to fully sample similar \hI\  
clouds in the M33 disk outskirts.

\section{General properties of the atomic neutral gas in M33}
\label{sec:h1line}

\subsection{HI channel maps, profile and mass}
Figure~\ref{fig:chlmap} presents selected channel maps of the combined DRAO+Arecibo datacube. 
The Galactic \hi\ emission does not appear in these channel maps. The foreground Galactic \hi\ is only detected at 
 $V_{\textsc{hel}} \ge -50$~\kms\ \citep[see also][]{Chemin2009} and does not contaminate the M33 gas emission. 
The variation of the orientation of the contours illustrates the perturbed \hi\ disk of M33.
 In particular, the contours of lower flux density do not draw a  
 V-shape typical of an unperturbed disk, but  are elongated 
 and twisted at their edges.  Also, the orientation of gas 
 at both ends of the  minor axis (heliocentric velocity $-178$ \kms) is almost perpendicular to that of the inner distribution. 
 These are the signatures of the  \hi\ warp of M33.

\begin{figure}[t!]
\includegraphics[width=\columnwidth]{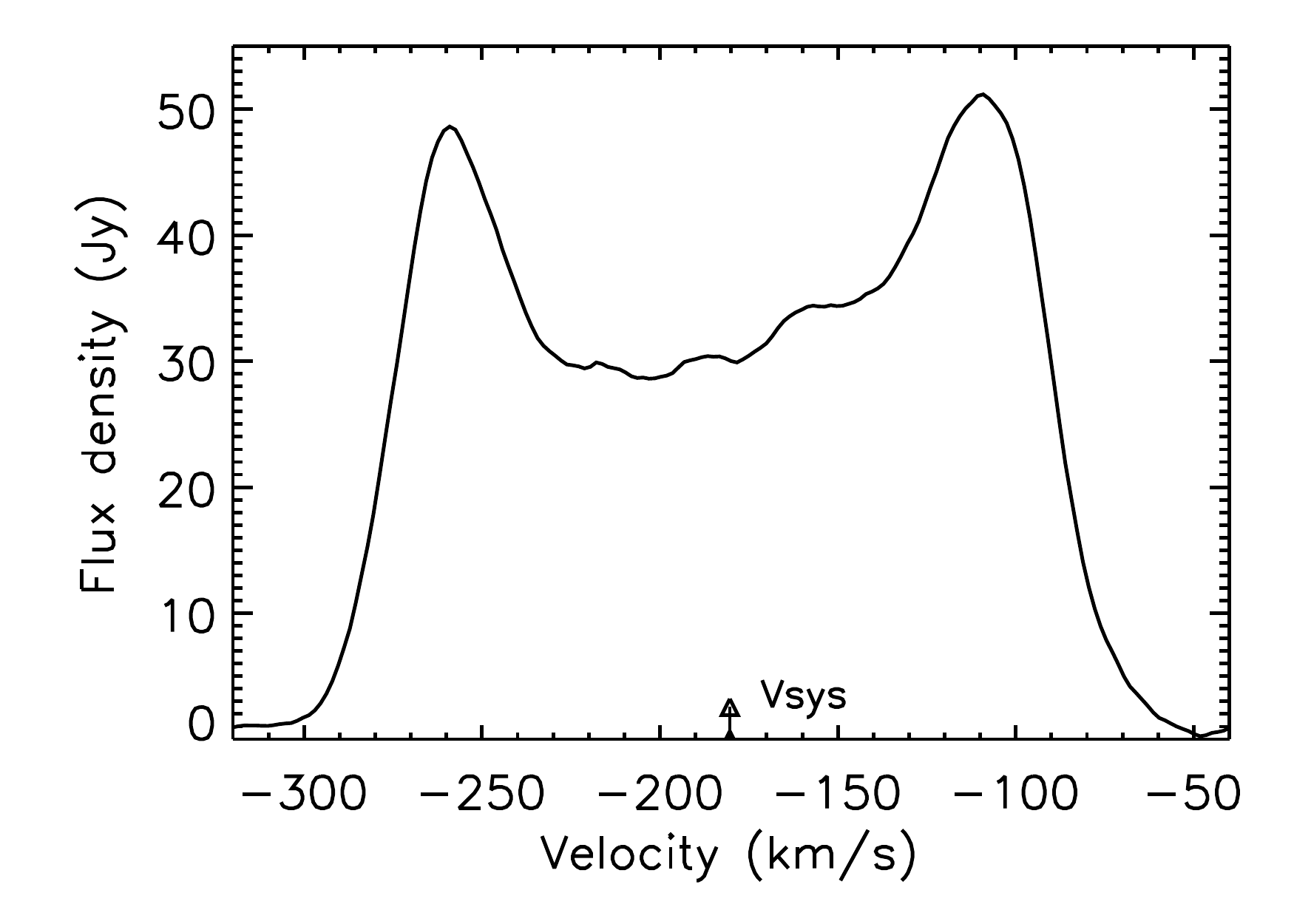}
\caption{Integrated \hi\ profile of M33.} 
\label{fig:globprof}
\end{figure}

The \hI\ integrated profile shown in Fig.~\ref{fig:globprof} is  asymmetric, with slightly more gas  in the receding Southern half (3\%\ relatively to the 
approaching Northern half).
The intensity weighted systemic velocity  is $-180.3\pm 2.3$~\kms, as derived from any channels with velocities $< - 50$ \kms. 
We define the maximum  flux of the global \hi\ profile as the average value of the two maxima.   
The width of the \hi\ profile at 50\% of the maximum flux is  $W_{50}=183$~\kms, with a corresponding 
mid-point velocity of $-180.4$~\kms, similar to the intensity weighted mean 
value. 
At the 20\% level, the velocity  width is $W_{20}=200$~\kms. The total \hI\ mass is  $1.95\pm 0.36 \, 10^9$ \msol. This value is in agreement with the mass found by \citet{Corbelli2014} from VLA-GBT data, and 
unsurprisingly with the one derived by \citet{Putman2009}.   
 The total neutral gas mass is about six times larger 
 than the molecular gas mass of $\sim 3.3 \times 10^8$ \msol\ 
 given in \citep{Gratier2010}. 

 \subsection{Anomalous velocity gas in M33}
  \label{sec:hilag}
\begin{figure}[t!]
  \includegraphics[width=\columnwidth]{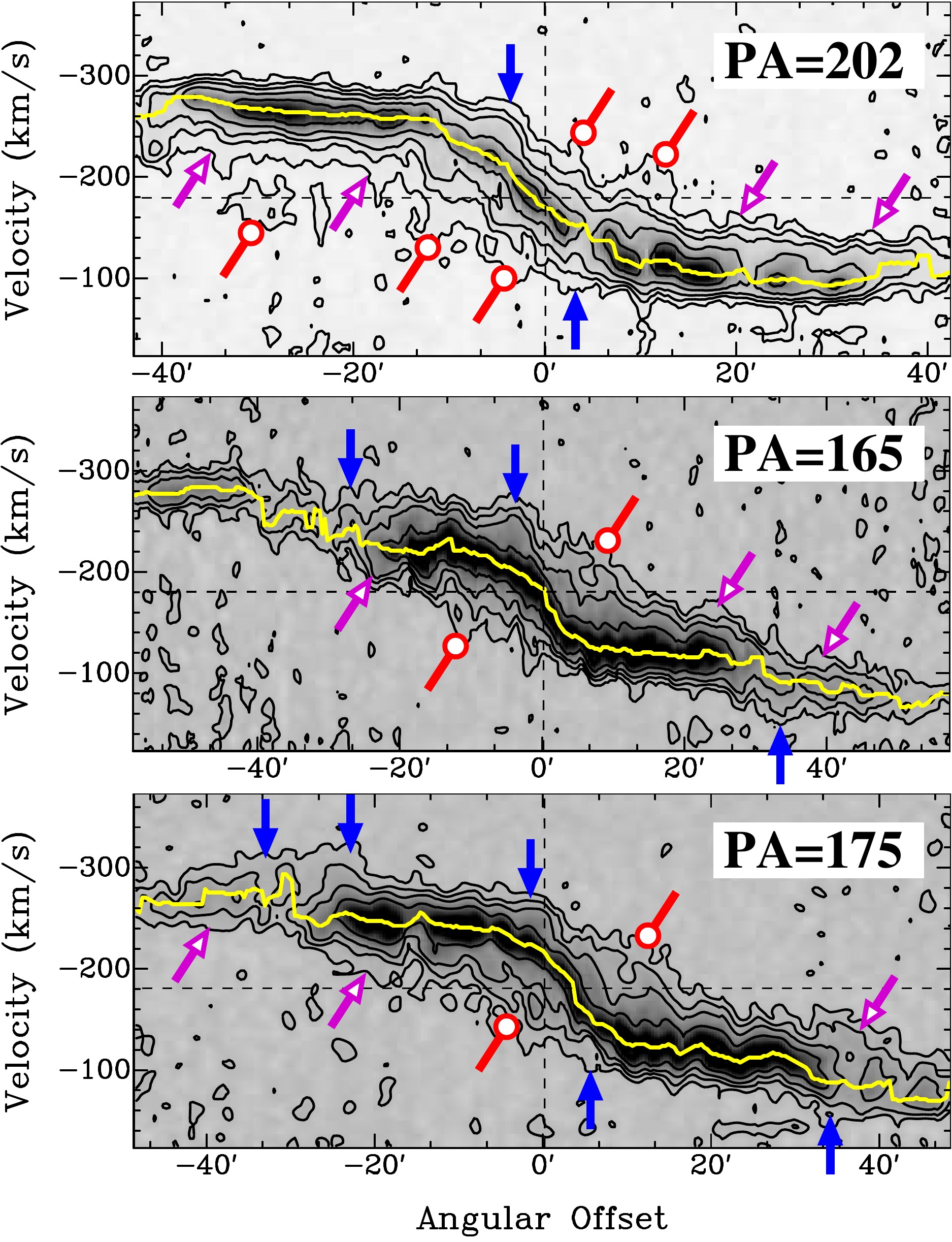}
 \caption{Position-velocity diagrams of M33. Datacube slices are made along the major axis of the inner
 disk (PA=202\degr, top panel), along the major axis of the outer disk (PA=165\degr, middle panel)  and 
 centered on the photometric centre of M33. The bottom panel is for PA=175\degr, centered  
 on  $(\alpha,\delta)_{\rm J2000} = (01{\rm h}33{\rm m}45.7{\rm s},+30\degr42'51'')$. 
 This explains the off-centered brightness distribution for this PV diagram.
 For each orientation, a yellow line is a cut 
 made in the \hi\ velocity field, showing the rotation of the high column density gas of M33. Contours  are 0.4, 1.2, 2.5, 6 and 15 K.  Arrows show the low density gas 
 lagging (open magenta pointers) and exceeding (filled blue pointers) the rotation of the 
 disk, while red circle pointers show the low density gas leaking in the forbidden velocity zones.}
 \label{fig:pvdiag} 
 \end{figure}

 Figure~\ref{fig:pvdiag}  presents position-velocity (PV) diagrams of the datacube, one 
 made along the major axis of the inner disk (position angle PA of 202\degr, see \S\ref{sec:kinparam}), one  
 along the major axis of the outer warped disk (PA = 165\degr), both of them
 being centered on the photometric centre, and another one along the direction PA=175\degr, but
 slightly off-centered from the photometric centre (bottom panel).  
 The center of this last PV diagram is  $(\alpha,\delta)_{\rm J2000} = (01{\rm h}33{\rm m}45.7{\rm s},+30\degr42'51'')$. 
 This slice orientation is chosen because it goes through 
 two regions of larger velocity dispersion at the NNW and SSE  (see \S\ref{sec:moments}). 
 The width of each PV slice is equivalent to two pixels ($\sim 45\arcsec$). 
 The yellow line in the diagrams traces a cut in the velocity field and highlights 
 nicely to which extent the high-brightness emission traces the rotation of the 
 disk.

 Along the direction $\rm PA=202\degr$, the high brightness gas is typical of a disk having slightly rising/constant 
 rotation curve (offsets $\le 30-35\arcmin$). 
 The brightness contours then exhibit   
 decreasing velocities at absolute offsets $> 35\arcmin$.  
 This apparent decrease occurs in the same region as 
 the one where the twist of the iso-brightness contours are observed in the channel maps, 
 as caused by the M33 warp.  
 Along the direction PA=165\degr,  
 the high brightness emission follows again 
 a slowly rising/flat rotational pattern.
 
 The most important result here is the detection of  
 a low brightness \hi\ component that has an extremely anomalous velocity behaviour with respect to the
  high-density gas. The contours of low density gas display ``beard-like'' features in the diagrams, 
  in reference to earlier works \citep[][and references therein]{Fraternali2001, Sancisi2001}, which means that  
  the contours are systematically stretched towards a line-of-sight velocity that deviates from those of contours of gas with the highest density.
 
\begin{figure}[t!]
  \includegraphics[width=\columnwidth]{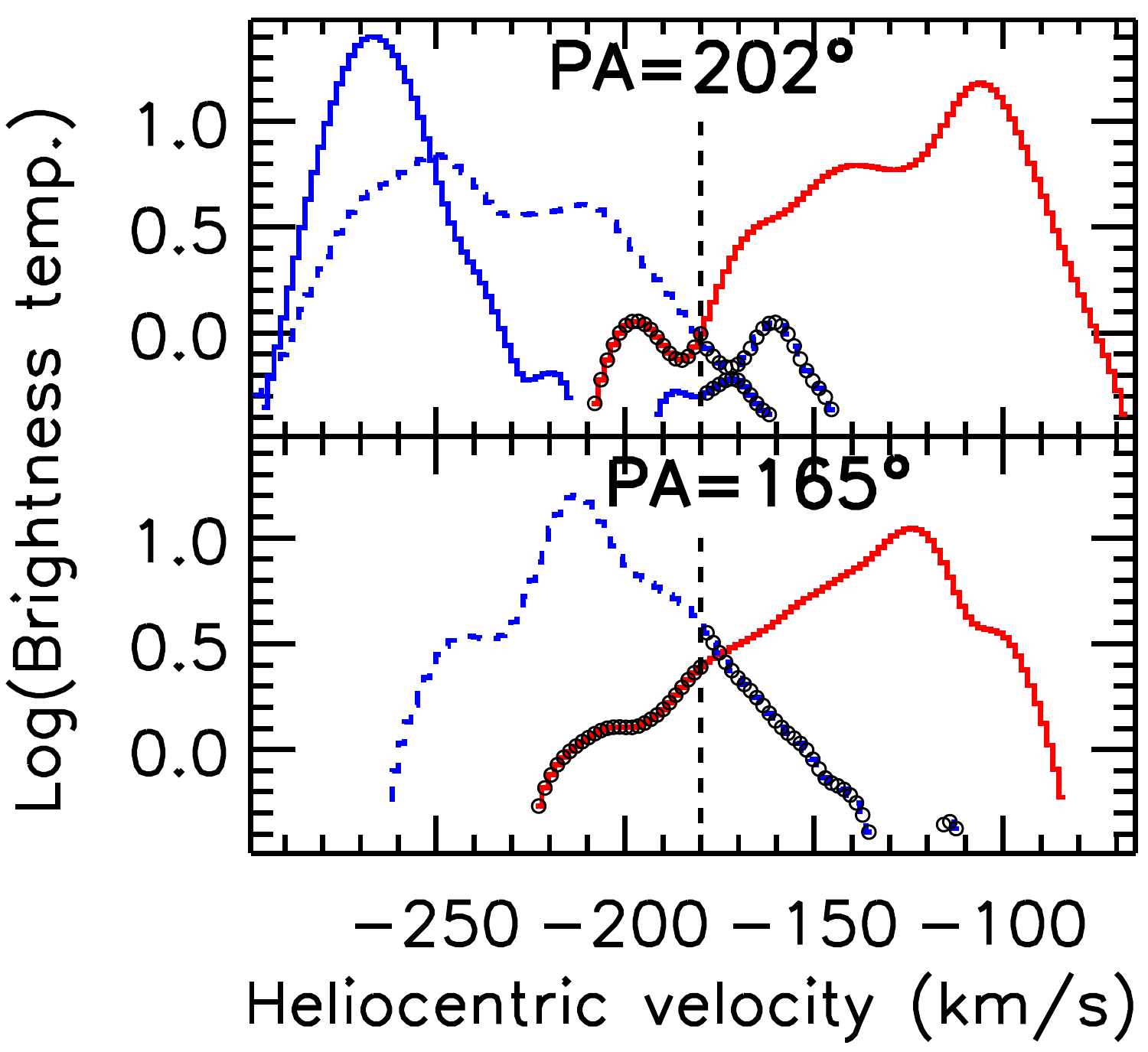}
 \caption{Five representative \hi\ profiles with a velocity component in the forbidden 
velocity zones from the PA=165\degr\ and PA=202\degr\ PV diagrams of Fig.~\ref{fig:pvdiag}. 
The coordinates of these spectra are given in Tab.~\ref{tab:coordgasFVZ}.
For each \hi\ profile, the velocity component in the forbidden zone is drawn as open circles. 
A dashed line is for offsets  $-11\arcmin$ and $-12\arcmin$, while solid lines are for other offsets. Blue and red lines
 indicate the approaching and receding disk sides, respectively. A vertical dashed line is the systemic velocity. 
 Temperatures are in Kelvin.}
 \label{fig:fvz} 
 \end{figure}

 First, the low brightness emission extends
 towards the systemic velocity of the system, as seen in the 
 top-left and bottom-right quadrants of all diagrams (open arrows). 
 Second, this low brightness gas  extends to higher velocities 
 than the high-brightness disk contours (filled arrows). This is more
 obvious for PA=165\degr\  and 175\degr\ than for PA=202\degr. In this PV diagram, only an absolute 
 offest $\sim 5\arcmin$ shows a high-velocity bump. 
 Third, the low brightness emission leaks in the forbidden velocity 
 zones, which makes it being in apparent counter-rotation with respect
  to the high-density gas. 
In our diagrams, the forbidden zones are the bottom-left and top-right quadrants. 
 A \hi\ component in the approaching disk half is in the forbidden velocity zone when its radial velocity is larger than the systemic 
 velocity (``receding component on the approaching side''). Conversely, a \hi\ component in the receding disk half 
 is in the forbidden velocity zone when its radial velocity is smaller than the systemic 
 velocity (``approaching component on the receding side'').  The forbidden velocity gas is detected 
at low offset and seems to make a link between the low and high velocity gas. Note also the isolated component
at 30\arcmin\ for PA=202\degr.
 We show in Fig.~\ref{fig:fvz} the gas component with forbidden velocity 
 by extracting \hi\ profiles at several positions along the PA=165\degr\ and 202\degr\ directions. Only channels with 
 $\rm T_B >5\sigma$ are shown for clarity. The equatorial coordinates of these spectra are given in Tab.~\ref{tab:coordgasFVZ}. 
 The  \hi\ gas in the forbidden zone is   the asymmetric tail of the  
 profiles, with velocities differing by up to 80-100\kms\ from the main peak. 
 \hi\ gas in the  PA=202\degr\ forbidden zone is also observed as distinct \hi\ peaks with velocities differing by 100 \kms\
 from the main peak (offsets$=-30\arcmin,+10\arcmin$).

\begin{table}
\centering
\caption{Coordinates of five representative spectra exhibiting a \hi\ velocity component in the forbidden 
velocity zones from the PV diagrams of Fig.~\ref{fig:pvdiag}.}
\begin{tabular}{lcc}
\hline
\hline
PV diagram &  Offset & Coordinates  (RA,~DEC) (J2000) \\
\hline
PA=165\degr & $+7\arcmin$ & 01$^{\textrm{h}}$33$^{\textrm{m}}$40.7$^{\textrm{s}}$,~+30$\degr$31$\arcmin$56$\arcsec$ \\
PA=165\degr & $-11\arcmin$ &01$^{\textrm{h}}$33$^{\textrm{m}}$22.0$^{\textrm{s}}$,~+30$\degr$49$\arcmin$46$\arcsec$ \\
PA=202\degr &  $+10\arcmin$ & 01$^{\textrm{h}}$33$^{\textrm{m}}$17.1$^{\textrm{s}}$,~+30$\degr$29$\arcmin$43$\arcsec$ \\
PA=202\degr &  $-12\arcmin$ &01$^{\textrm{h}}$33$^{\textrm{m}}$52.6$^{\textrm{s}}$,~+30$\degr$50$\arcmin$09$\arcsec$ \\
PA=202\degr &  $-30\arcmin$ & 01$^{\textrm{h}}$34$^{\textrm{m}}$26.7$^{\textrm{s}}$,~+31$\degr$07$\arcmin$49$\arcsec$ \\
\hline
\end{tabular}
\label{tab:coordgasFVZ}
\end{table}

  The first anomalous component  is reminiscent of the slow rotation extraplanar 
   \hi\ layer  seen  in other galaxies  \citep[e.g.  NGC2403, NGC891, NGC253,][]{Fraternali2002, Oosterloo2007,
   Lucero2015}.
 The second component  reminds the
   population  of low mass, high-velocity clouds in the halo of our Galaxy \citep{Wakker1997} or M31 \citep{Westmeier2008}.  
     \citet{Putman2009} already evidenced high-velocity gas around M33, 
   but at larger distances than the new one detected here, well outside the field-of-view of our observtions.
   These authors proposed a scenario with extraplanar gas falling onto M33 after a tidal event with M31. 
   As for the third component, it would not be the first time that forbidden velocity  gas
   is reported in  nearby galaxies \citep[NGC2403,][]{Fraternali2001, Fraternali2002}. Interestingly, these authors reported 
   that in NGC2403 the low angular momentum and the forbidden velocity components seem to form a coherent structure. 
   M33 is thus similar to NGC2403 in that aspect.
    
    The detailed derivation of the distribution, kinematics and mass of  the three  anomalous components 
    is beyond the scope of this
    article and will be presented in a future paper (Chemin et al., in preparation).  This work will require a more appropriate
    processing of the  datacube into several components than the single moment maps analysis we did here, maybe like in  
    \citet{Fraternali2002} with NGC2403, who fitted and subtracted to the datacube a Gaussian profile 
    centred on the highest \hi\ peak, or in \citet{Chemin2012},  who fitted multiple Gaussians peaks to the current M33 dataset.

%% file: Velocity.tex
\subsection{Moment maps}
\label{sec:moments}  

\begin{figure*}[t]
\includegraphics[width=0.5\textwidth]{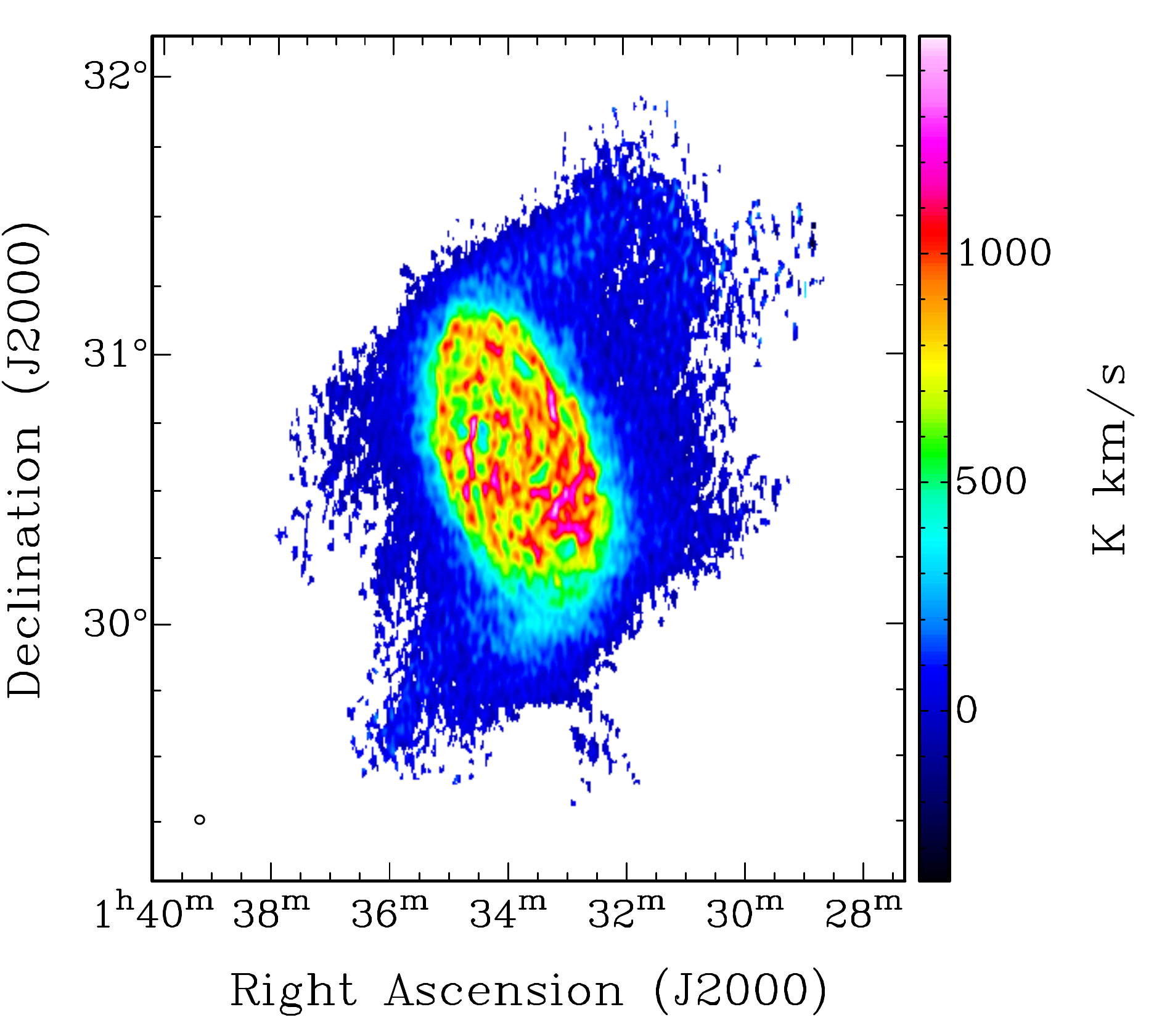}\includegraphics[width=0.5\textwidth]{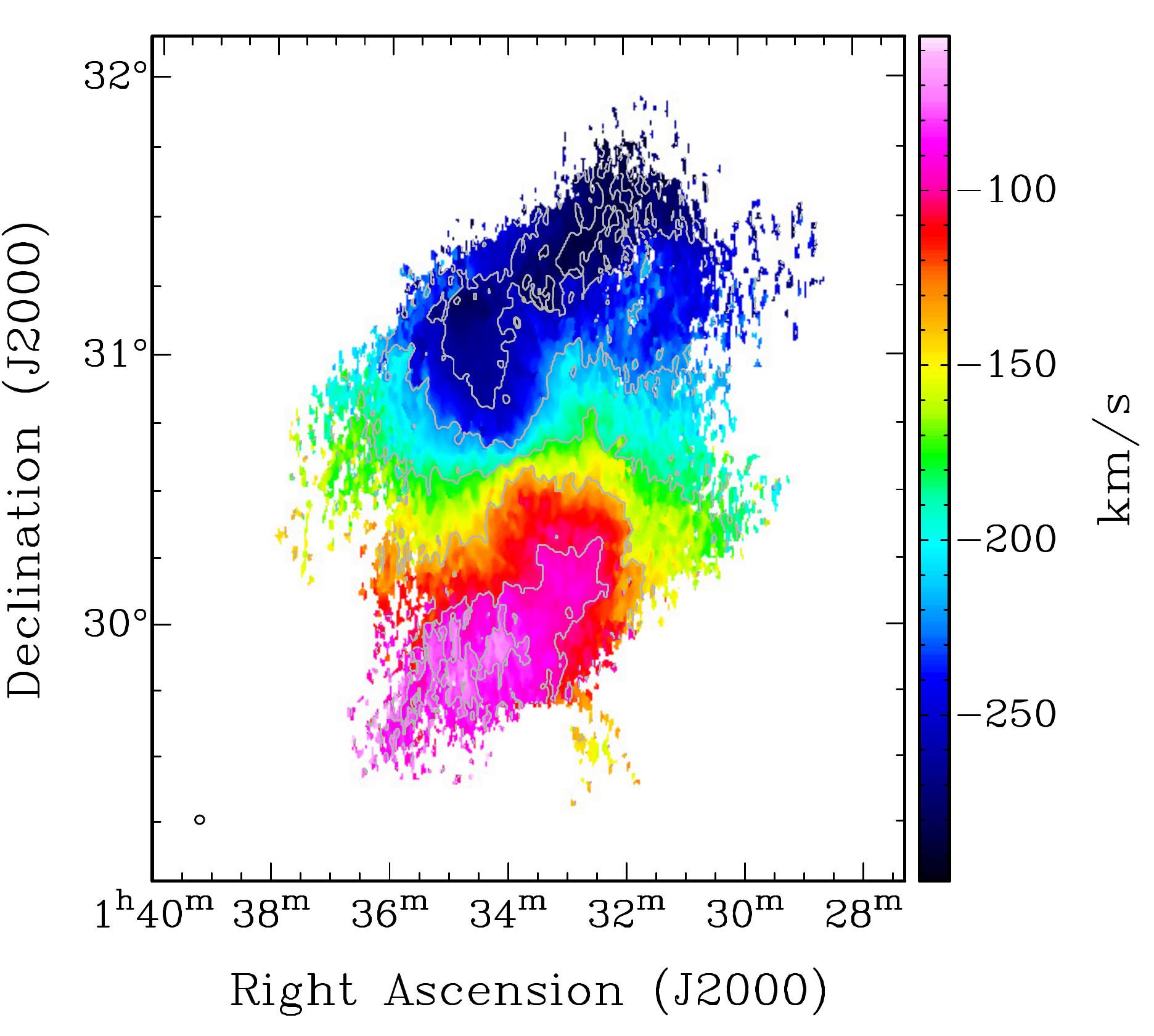}
 \caption{\hi\ integrated emission map  and velocity field  of Messier 33 (left and 
 right panels, respectively). 
 The velocity contours are  $-280$, $-260$, $-220$, $-180$, $-140$, $-100$, 
  and $-80$ \kms.  The circle to the bottom left of each panel represents the 2\arcmin\ resolution.}
 \label{fig:velomap} 
 \end{figure*}

\begin{figure}[t]
\includegraphics[width=0.5\textwidth]{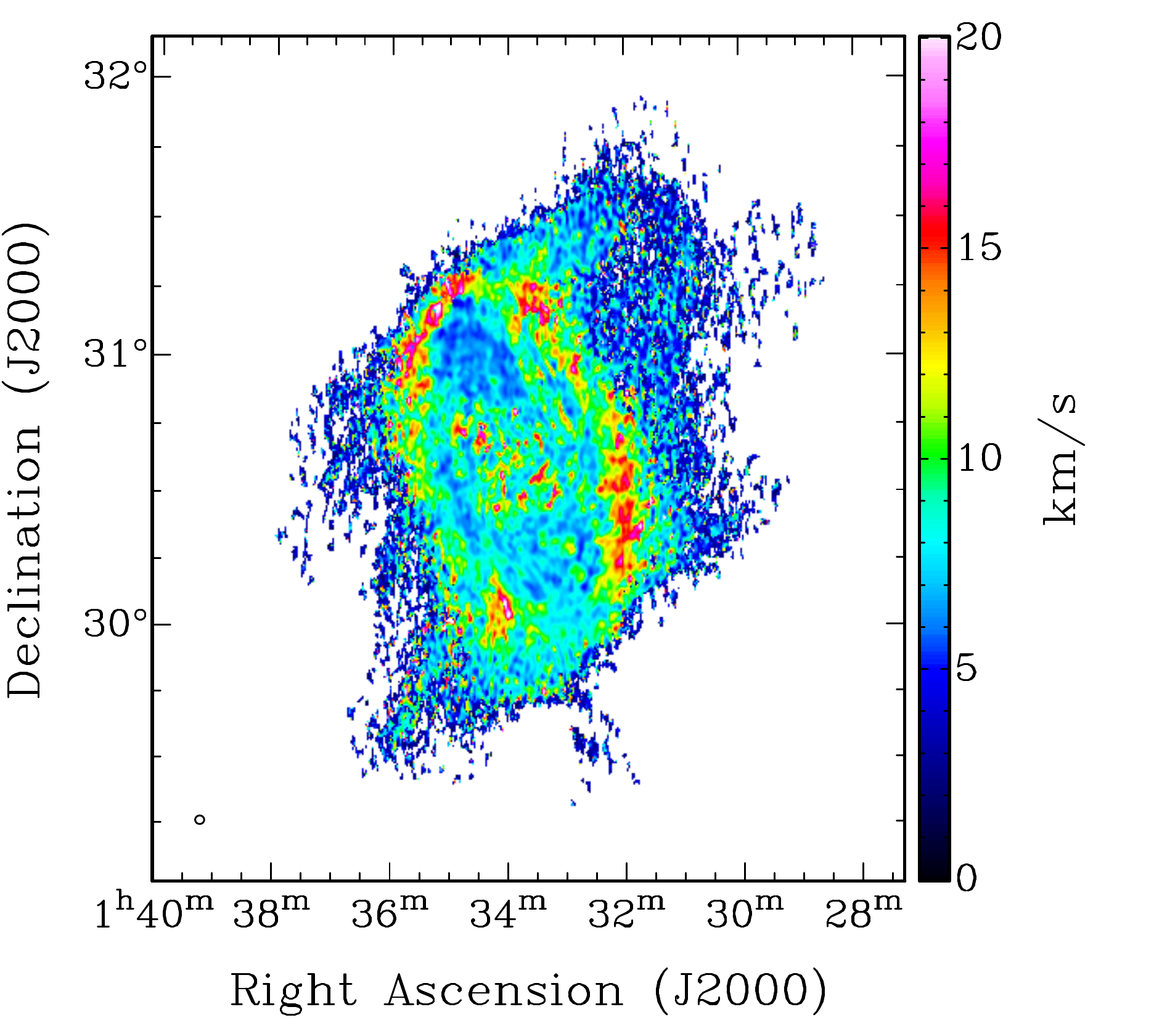}
 \caption{\hi\ velocity dispersion field of Messier 33.  The circle to the bottom left of each panel represents the 2\arcmin\ resolution.}
 \label{fig:velomap2} 
 \end{figure}

The task  \textsc{moments}  in \textsc{gipsy}  has been used to compute the moment maps  using the data cube smoothed at 
$120\arcsec \times120\arcsec$. Figures~\ref{fig:velomap} and~\ref{fig:velomap2} 
show the maps of the $0^{\rm th}$ moment (integrated emission), $1^{\rm st}$ (velocity field) 
and $2^{\rm nd}$ (velocity dispersion). 

The \hi\ emission map shows that the high surface density gas is contained within the stellar disk ($R \lesssim 30-35\arcmin$,
or $\lesssim 8$ kpc). 
It displays a multiple spiral arms pattern that coincides well with the spiral structure evidenced in the  \ha\ disk (Kam15) and  
molecular gas disk \citep{Druard2014}. 
Beyond the stellar disk,  the gas column density sharply decreases, reaching $\sim 5 \times 10^{18}$ cm$^{-2}$ in the outermost regions.
Gaseous tails and arcs are observed to the North-West and South-East of the disk as part of the \hi\ warp.
The  SE feature is less extended than the NW arc-like structure,  
which roughly points in the direction of M31. 
 These perturbations could be signatures of the past interaction between the two galaxies.
Note also that M33 is surrounded by a prominent stellar structure that provides additional evidence of an  
encounter with M31 \citep{McConnachie2010}, or at least of recent gravitational interaction. 
This stellar structure extends $\sim 2$\degr\ (30 kpc in projection) to the North-West towards M31, nearly three times farther 
 out than the size of the M33 stellar disk, and thus farther than the \hi\ gas in the warp. 

\begin{figure}[t]
\includegraphics[width=\columnwidth]{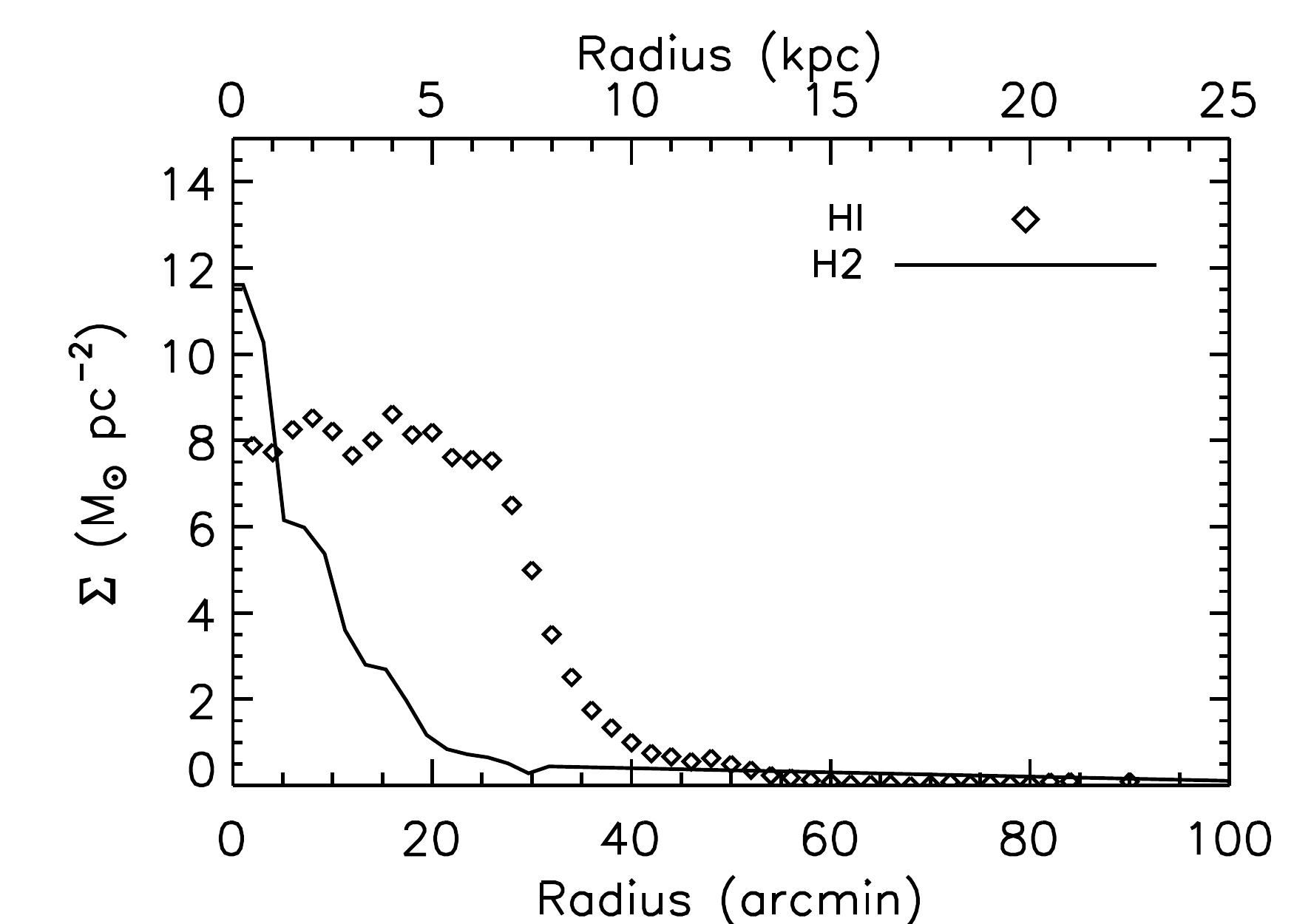}
\caption{\hi\ mass surface density profile of M33 (symbols). The molecular gas mass surface density (solid line) is from \citet{Druard2014}.} 
\label{fig:radprof}
\end{figure}

Our \hi\ map is in very good agreement with  atomic gas distributions seen 
elsewhere \citep{Corbelli2014}.  
The \hi\ mass surface density profile derived from the column density map with the adopted kinematical parameters given below 
is shown in Fig.~\ref{fig:radprof}. The \hdeux\ mass surface density from \citet{Druard2014} is also shown.
 Naturally, \hdeux\ is much more concentrated  than the atomic gas. Both gas components reach
  surface densities $\sim$10 \msol\ $\rm pc^{-2}$ (8 for \hi\ and 12 for \hdeux) in the inner disk regions.

The velocity field is quite regular in the inner regions ($R \lesssim 8$ kpc). 
Beyond that radius, in the regions where the gas distribution is perturbed, the velocity field is irregular.
 This is apparent in the prominent twist of the velocity contours, whose feature was also evidenced in the channel maps (Sect.~\ref{sec:h1line}). 
Note also an outer South-Western cloud or extension, illustrated 
 by yellow and orange colours in the velocity map, that rotates slower than 
gas inside the disk at similar radius.
This extension has already been discussed by \cite{Putman2009}  \citep[see also][]{Grossi2008}. 
 
\begin{figure*}[t]
\includegraphics[width=\textwidth]{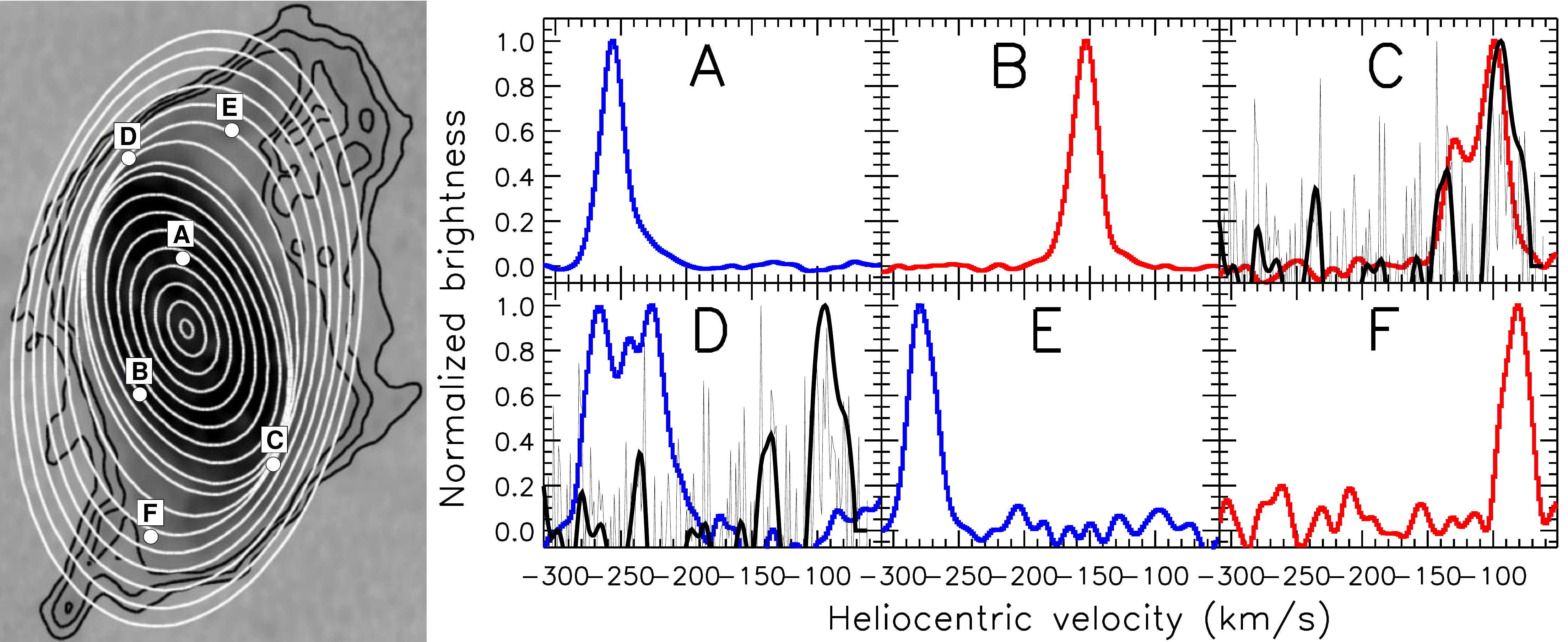}
 \caption{Geometry of the tilted rings from the kinematical modeling (left panel) and 
 \hi\ profiles selected at six locations in the galaxy (right panels). 
 The grey-scale image is the integrated \hi\ emission with contours highlighting low column densities at the outskirts of the \hi\ disk.  
 The ellipses are separated by 240\arcsec\ from each other. 
 Thick blue and red lines are profiles from the DRAO datacube for the approaching and receding disk 
 sides, respectively. Black lines at positions C and D are profiles  from the 12\arcsec\ resolution VLA 
 datacube of \citet{Gratier2010}. A thin black line is for the original spectral
  resolution of VLA profiles, while 
 a thick black line is for VLA profiles
  convolved at the same spectral resolution as the DRAO data. }
  \label{fig:warp}
 \end{figure*}

 In the velocity dispersion map, an incomplete ring of larger dispersion
 is observed in the transition between the high column density inner disk 
  and the low column density outer disk.  In this structure, profiles exhibit multiple \hi\ peaks  
  or wider single \hi\ peaks \citep[see also][]{Chemin2012}. These features are 
  likely due to the crowding of gas orbits because of the disk warping (see \S\ref{sec:warp} and Fig.~\ref{fig:warp}).

 \begin{figure}[t!]
 \centering
\includegraphics[width=\columnwidth]{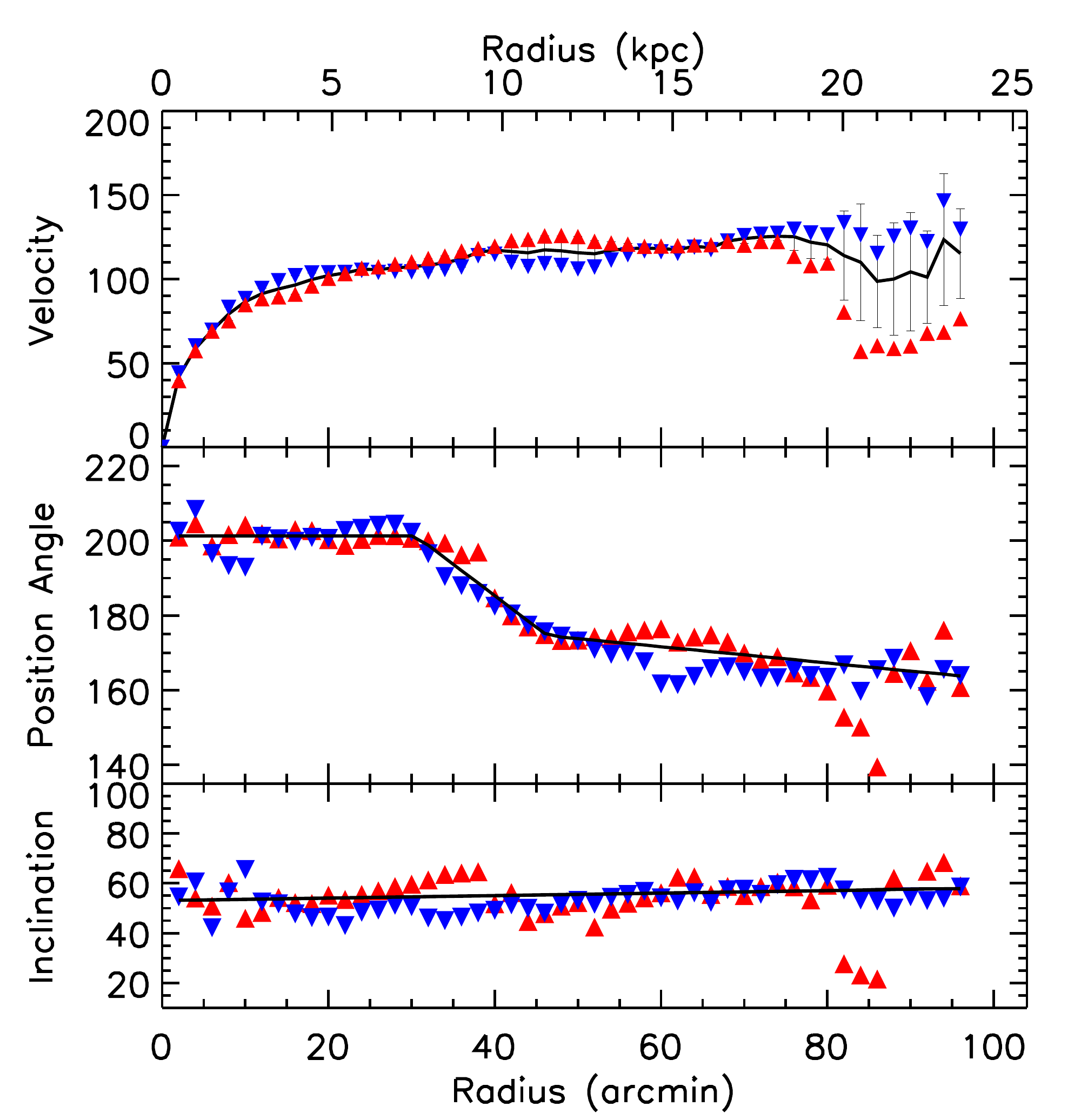}
 \caption{Results of the tilted-ring model of the \hi\ velocity field of M33.  
 The top panel shows the rotation curve (in \kms), the middle panel the major axis position angle (in \degr) and the bottom panel the inclination (in \degr). 
 Red downward triangles are the results for the receding 
 side, blue upward triangles those for the approaching side. 
 Solid lines are the adopted profiles from both sides.}
 \label{fig:restiltedringmodel}
 \end{figure}

\subsection{Derivation of the kinematical parameters}
\label{sec:kinparam}  
The task \textsc{rotcur} in \textsc{gipsy} was used to derive the kinematical parameters and the rotation curve. 
 Assuming negligible radial motions,  the tilted-ring model fits the expression
\begin{equation}
{ \rm V_{obs} = V_{sys} +V_{rot} } \cos\theta \sin i  
\label{eq:vobs_mo}
\end{equation} 
to the \hi\ velocity field, where  $\theta$ is the azimuthal angle in the plane of the galaxy and 
$i$ the inclination. The position angle of the kinematical major axis is defined as the counterclockwise angle in the plane of the sky  
from the North to the receding side semi-major axis.  The angle $\theta$ is measured relatively to the semi-major axis.
The South-Western Cloud has  been masked for the derivation of the parameters and rotation curve.

In a first run, the position of the rotation centre
 and the systemic velocity $\rm V_{sys}$  were let free to vary, using fixed kinematical 
 inclination and major-axis PA   at the optical values (52\degr\ and 202\degr, respectively). 
 We find a systemic velocity of $\rm V_{sys} \sim -183$ \kms, and a centre at $(\alpha,\delta)_{\rm J2000}=(01{\rm h}33{\rm
 m}50.9{\rm s},+30\degr40'20'')$.
Since the positional difference with respect to the photometric centre is much less than the beam size, 
we decided to fix the kinematical centre at the position of the photometric centre. 
Moreover, because the inferred systemic velocity is in agreement with the  
mean value  determined from the integrated profile (\S\ref{sec:moments}), we adopted 
$\rm V_{sys} =-180$~\kms, as given by the profile (both intensity weighted mean and mid-point velocity). 
A second run allowed us to fit the inclination and  major-axis PA using fixed centre and systemic velocity  given by the adopted values. 
The results of the tilted ring model are shown in Fig.~\ref{fig:restiltedringmodel} and the adopted 
inclination and position angle are listed in Tab.~\ref{tab:rvHI}. 

 \begin{figure}[b]
 \includegraphics[width=\columnwidth]{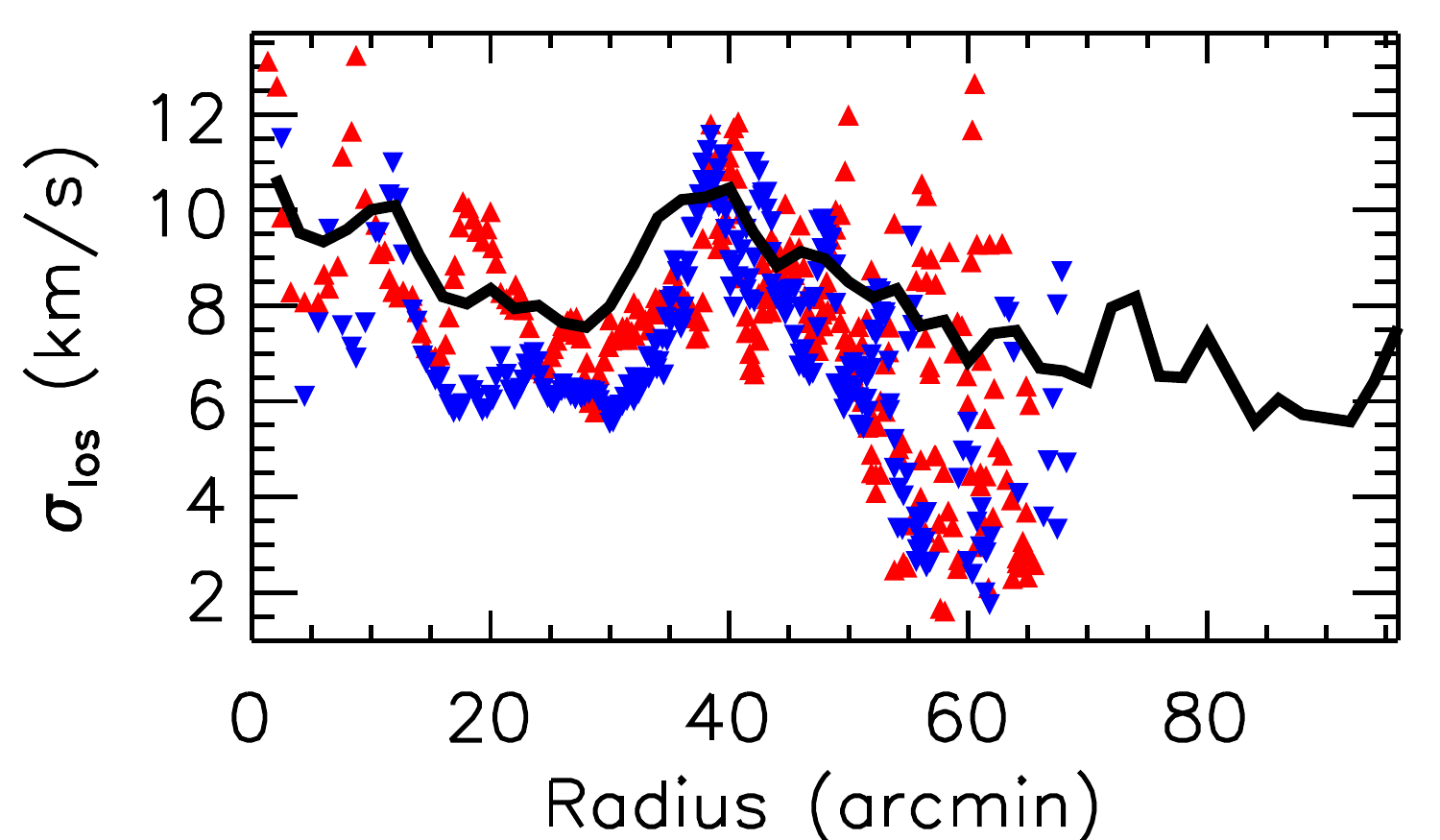}
  \caption{Azimuthally averaged line-of-sight velocity  dispersion profile (solid line). Blue and red symbols are the observed dispersions 
  along the semi-major axis of the dispersion field for the approaching and receding disk halves, respectively.}
  \label{fig:siglos}
\end{figure}

\subsection{The HI rotation curve and velocity dispersion profile}
\label{sec:RC} 

The \hi\ rotation curve (Fig.~\ref{fig:restiltedringmodel}) 
is then derived  with fixed values for the kinematical centre and systemic velocity, 
and the adopted models of inclination and position angle (solid lines in the middle and bottom panels). 
Also shown are the obtained \hi\ rotation curves for the approaching ($\rm V_a$) and receding sides 
($\rm V_r$) fitted 
separately with similar kinematical parameters.
Differences between $\rm V_a$  and $\rm V_r$  are mostly smaller than 10 \kms. The largest differences, up to 60 \kms, occur at $R> 80\arcmin$ in the  perturbed 
NW and SE structures from the warp.

The \hi\ rotation curve is reported in Tab.~\ref{tab:rvHI}. The total velocity uncertainty  $\rm \Delta V_{rot}$  is defined by
  $\rm \Delta V^2_{rot}=\epsilon^2 + |(V_a-V_r)/2|^2$ with $\epsilon$ being the formal RMS error for the model 
  with both sides  fitted simultaneously (solid line in the 
top panel of Fig.~\ref{fig:restiltedringmodel}). The dominant error of the total uncertainty is the velocity difference between the two disk sides.    
Because of that definition, the errors
appear larger than in many other studies. However, we prefer being conservative 
because we model the rotation curve with axisymmetric components (Sect.~\ref{sec:massmod}) and consider 
that our definition is more representative of the true asymmetry in the observed kinematics 
and of uncertainties when doing mass models.

\begin{table*}
\begin{center}
\caption{Results of the tilted-ring model of the \hi\ velocity field of M33.}  
\label{tab:rvHI}
\begin{tabular}{cccccc|cccccc}
\hline\hline 
 Radius & Radius &   $\rm V_{rot}$ & $\rm \Delta V_{rot}$ &  $i$ & PA &  Radius  &  Radius &  $\rm V_{rot}$ &  $\rm \Delta V_{rot}$  & $i$ &  PA\\ 
 (\arcmin) &  (kpc) &  (\kms)  &  (\kms) &  (\degr) & (\degr) &  (\arcmin) &  (kpc) &  (\kms) &  (\kms) &  (\degr) & (\degr)\\
\hline
     2  &   0.5   &   42.0   &	2.4   &   53.2   &   201.3   &   50  &   12.2	&   115.7   &	9.6   &   55.6   &   173.8 \\
     4  &   1.0   &   58.8   &	1.5   &   53.3   &   201.3   &   52  &   12.7	&   115.1   &	7.7   &   55.7   &   173.4 \\
     6  &   1.5   &   69.4   &	0.4   &   53.4   &   201.3   &   54  &   13.2	&   117.1   &	5.1   &   55.8   &   172.9 \\
     8  &   2.0   &   79.3   &	4.0   &   53.5   &   201.3   &   56  &   13.7	&   118.2   &	3.2   &   55.9   &   172.5 \\
    10  &   2.4   &   86.7   &   1.8   &   53.6   &   201.3   &   58 &   14.2   &   118.4   &   1.4   &   56.0   &   172.1 \\
    12  &   2.9   &   91.4   &   3.1   &   53.7   &   201.3   &   60 &   14.7   &   118.2   &   1.8   &   56.1   &   171.6 \\
    14  &   3.4   &   94.2   &   4.8   &   53.8   &   201.3   &   62 &   15.1   &   117.5   &   2.4   &   56.2   &   171.2 \\
    16  &   3.9   &   96.5   &   5.5   &   53.9   &   201.3   &   64 &   15.6   &   119.6   &   0.8   &   56.3   &   170.8 \\
    18  &   4.4   &   99.8   &   3.9   &   54.0   &   201.3   &   66 &   16.1   &   118.6   &   1.5   &   56.4   &   170.3 \\
    20  &   4.9   &   102.1   &   1.7	&   54.1   &   201.3   &   68  &   16.6   &   122.6   &   0.5	&   56.5   &   169.9 \\
    22  &   5.4   &   103.6   &   0.4	&   54.2   &   201.3   &   70  &   17.1   &   124.1   &   2.9	&   56.6   &   169.5 \\
    24  &   5.9   &   105.9   &   0.7	&   54.3   &   201.3   &   72  &   17.6   &   125.0   &   2.2	&   56.7   &   169.0 \\
    26  &   6.4   &   105.7   &   1.7	&   54.4   &   201.3   &   74  &   18.1   &   125.5   &   2.5	&   56.8   &   168.6 \\
    28  &   6.8   &   106.8   &   2.2	&   54.5   &   201.3   &   76  &   18.6   &   125.2   &   8.1	&   56.9   &   168.2 \\
    30  &   7.3   &   107.3   &   3.0	&   54.6   &   201.3   &   78  &   19.1   &   122.0   &   9.8	&   57.0   &   167.7 \\
    32  &   7.8   &   108.3   &   4.0	&   54.7   &   198.8   &   80  &   19.5   &   120.4   &   8.5	&   57.1   &   167.3 \\
    34  &   8.3   &   109.7   &   4.0	&   54.8   &   195.4   &   82  &   20.0   &   114.0   &   26.6   &   57.2   &	166.8 \\
    36  &   8.8   &   112.0   &   4.8	&   54.9   &   192.0   &   84  &   20.5   &   110.0   &   34.6   &   57.3   &	166.4 \\
    38  &   9.3   &   116.1   &   2.2	&   55.0   &   188.7   &   86  &   21.0   &   98.7   &   27.4	&   57.5   &   166.0 \\
    40  &   9.8   &   117.2   &   2.5	&   55.1   &   185.3   &   88  &   21.5   &   100.1   &   33.4   &   57.6   &	165.5 \\
    42  &   10.3   &   116.5   &   6.5   &   55.2   &	181.9	&  90 &   22.0   &   104.3   &   35.2	&   57.7    &	165.1 \\
    44  &   10.8   &   115.7   &   8.1   &   55.3   &	178.5	&  92 &   22.5   &   101.2   &   27.4	&   57.8    &	164.7 \\
    46  &   11.2   &   117.4   &   8.2   &   55.4   &	175.2	&   94   &   23.0   &   123.5   &   39.1   &   57.8	&   164.3 \\
    48  &   11.7   &   116.8   &   8.9   &   55.5   &	174.2	&   96   &   23.5   &   115.3   &   26.7   &   57.9	&   163.8 \\
\hline 
\end{tabular} 
\end{center}
Comments: $i$ and PA are the adopted \hi\ inclination and major axis position angle and $\rm V_{rot}$ the resulting \hi\ rotation curve
 with associated velocity uncertainties (see text for details).
\end{table*}

Figure~\ref{fig:siglos} shows the line-of-sigth dispersion profile of M33, derived by azimuthally averaging the dispersion field using the adopted inclination
 and position angle profiles. Table~\ref{tab:siglossurfdens} of
 Appendix~\ref{sec:siglossurfdens} lists the mean dispersion. On average, the dispersion is $\sim 9$ \kms\ in the central regions ($R<30\arcmin$). The signature of the 
 ring of higher dispersion is observed as a rise of 3 \kms\ ($R=35-40\arcmin$). The dispersion then continuously decreases at larger radius. 
 Cuts made along the semi-major axis of the approaching and receding sides in the dispersion map   show significant differences inside $R=30\arcmin$ 
 and beyond $R=50\arcmin$. Moreover, the mean dispersion is larger than values along the semi-major axes.  
 Such differences demonstrate the intrinsic asymmetry of the dispersion field, similarly to the disturbed velocity field, as caused by any of the spiral, warp, and 
 lopsidedness perturbations (see \S\ref{sec:warp} and~\ref{sec:asym}). 

\begin{figure}[t]
 \includegraphics[width=\columnwidth]{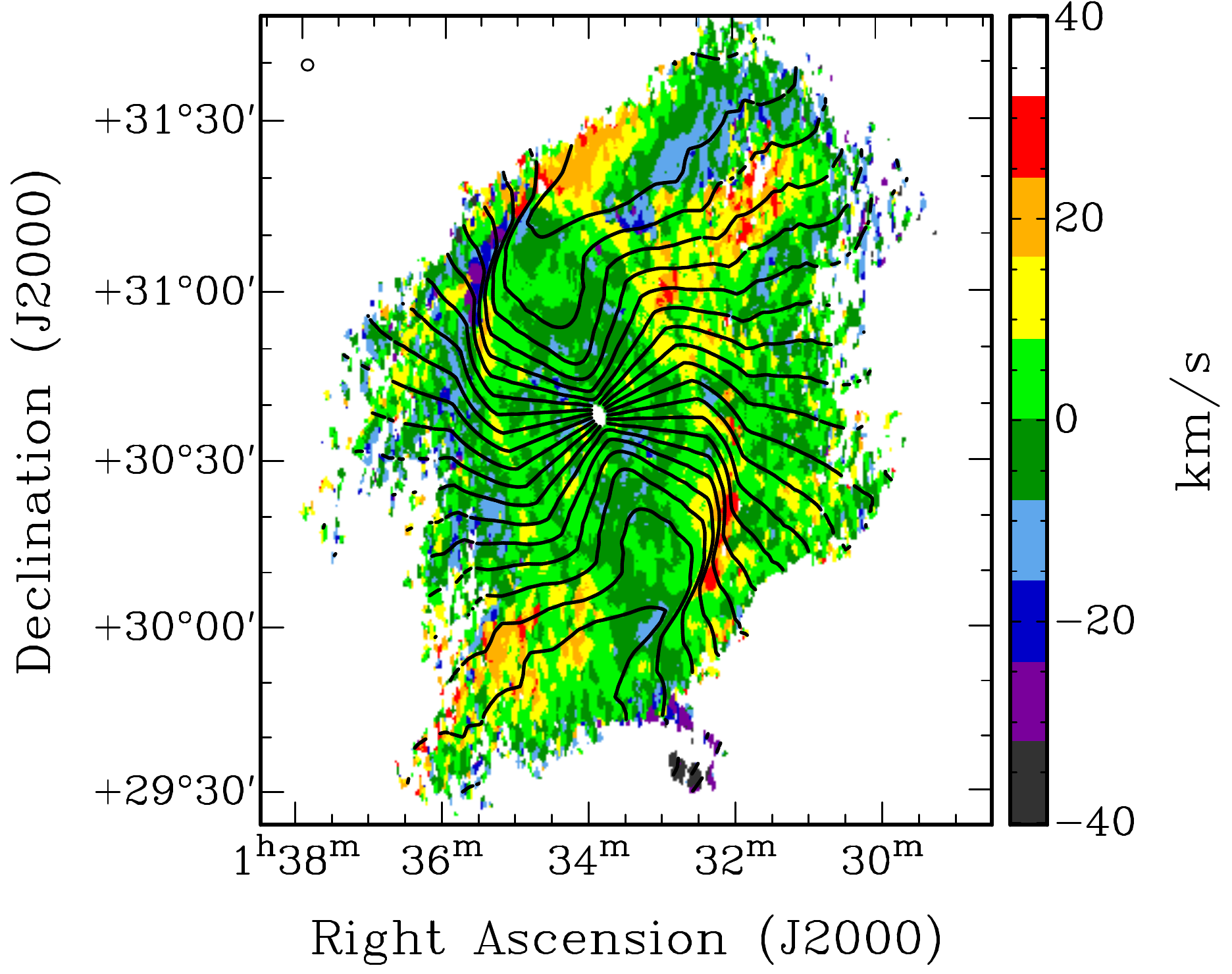}
  \caption{M33 residual velocity field (observation minus model). The model velocity field is represented by the contours, from -90 to -280 \kms\ by step of 10 \kms. 
   The circle to the upper left corner represents the 2\arcmin\ resolution.}
  \label{fig:model}
\end{figure}

\subsection{The HI warp of M33}
\label{sec:warp}  

The warping of M33 is mainly observed as a significant kinematical twist of the major axis position angle, 
starting by a sharp decrease from 
$\sim 202\degr$ to $\sim 170\degr$ within $7<R<11$ kpc, and then a smooth variation down to $165\degr$ at large radius. 
A kinematical tilt is also evidenced, though less spectacular, 
as a small rise of inclination ($\sim 5\degr$ from the centre to the disk outskirts). 

The  geometry of the \hi\ orbits implied by the axisymmetric tilted-ring model is shown in Fig.~\ref{fig:warp}. 
Within the  transition region of significant PA twist, the modeled rings are very close to each other.  
Gas clouds are likely colliding at these radii. 
This statement is confirmed by the observation of 
 double \hi\ peaks at e.g. locations C and D, or more generally by 
 the incomplete ring-like structure of larger velocity dispersion (Fig~\ref{fig:velomap}).  
 The change of inclination is so insignificant here that a configuration where 
 the line-of-sight crosses the disk more than once is hardly possible. 
 Moreover, we verified that the wide and double \hi\ peaks are not caused by the ``low'' resolution of the DRAO data. Indeed, first the 
 velocity gradient is not important in this region because the rotation curve is almost flat. Then, 
 such peculiar \hi\ profiles are also observed in the 12\arcsec\ VLA datacube of \citet{Gratier2010}, 
 as seen at position C (the location D is at the periphery of the VLA field-of-view, the corresponding 
 profile is only made of noise).  
 
 It is worthwhile to mention that not all wider profiles  
 can be explained by our idealized model in the transition region because some of them are also observed 
 elsewhere than at the location of crowded rings. This is seen at angular offsets $\sim 30-35\arcmin$ 
  in the PA=175\degr\ PV diagram of Fig.~\ref{fig:pvdiag}, which offsets correspond to  
  $(\alpha,\delta)_{\rm J2000}=(01{\rm h}33{\rm m}57.6{\rm s},+30\degr05'02'')$ and 
  $(01{\rm h}33{\rm m}20.2{\rm s},+31\degr12'30'')$. 
  Here, these wider profiles coincide with holes in the gas distribution and trace  
   high- and low-velocity faint gas (\S\ref{sec:hilag}). 
  
  Elsewhere than in the transition region, the inferred model geometry is regular
 and wide profiles are not observed (positions A and B in the inner disk, positions E and F in the outer disk).
 
Interestingly, the PA twist model implies that the outermost ellipses have their major axis 
aligned with the direction M33-M31. 
This could be a consequence of the gravitational interaction between the two galaxies.
In this scenario, the interaction stretched the outer \hi\ disk in the direction of M31, 
 and that perturbation is acting down to
a radius of $R \sim 7$ kpc. Circular orbits are excluded in these outer regions
as circularity cannot create precessed and closely grouped rings such as those implied by the projected model.  
Gas orbits are likely elongated and lopsided beyond that radius.

The model velocity map based on the \hi\ RC allows us to make a residual velocity map, defined here as the observation minus the model  velocity field (Fig.~\ref{fig:model}). 
No systematic velocity asymmetry is found, implying the goodness of the adopted mass centre and systemic velocity. 
 Significant residuals are tightly linked to the warp perturbation described above.
The lowest residuals occur in the limit of the unperturbed inner \hi\ disk, at the level of a few \kms. 
 The largest residuals are observed in the outer disk, starting from the transition 
 zone in which the major axis PA varies abruptly, up to $\sim 30$ \kms\ (absolute value). 
 The ring of higher velocity dispersion (Fig.~\ref{fig:velomap}) thus coincides with large residuals. 
 Note also the significant residual for the SW clump ($> 30$ \kms, absolute value), confirming that it does not rotate similarly as the disk.
 One can notice finally that observing larger residuals in the NW and SE \hi\ extensions confirms the 
 existence of asymmetric gas orbits at the disk periphery.

\subsection{Asymmetric and non-circular motions in M33}
\label{sec:asym}

We can assess further the perturbations in the \hi\ velocity field of M33 by expanding the standard  
 model of Eq.~\ref{eq:vobs_mo} to higher order cosine and sine terms. \citet{Franx1994} and \citet{Schoenmakers1997} were among the first to 
 propose Fourier analyses of velocity fields to estimate asymmetries caused by, e.g.,
  oval distortions, lopsided potentials, spiral arms or warps. 
 These authors argued that a perturbing mode of the gravitational potential of order $m$  generates, 
 to the first order, $k=m-1$ and $k=m+1$ Fourier components in velocity fields. 
 
  The following Fourier series model was thus fitted to the  \hi\ velocity field:  
\begin{equation}
  {\rm V_{obs}} =   c_0 +  \sum_{\rm k=1} \left( c_{k}\cos k\theta + s_{k}\sin k\theta \right) \sin i  
  \label{eq:fouriervf}
\end{equation}
where $c_k$ and $s_k$ are the velocity coefficients of harmonic order $k$ ($k$ is an integer). 
The $c_0$ coefficient is the systemic velocity $\rm V_{sys}$ of Eq.~\ref{eq:vobs_mo}, the $c_1$ coefficient is the rotation curve, the $s_1$ term is 
the noncircular radial velocity, while higher order terms constrain deviations from axisymmetry of $c_1$ and $s_1$.  

\begin{figure}[t]
\begin{centering}
\includegraphics[width=0.9\columnwidth]{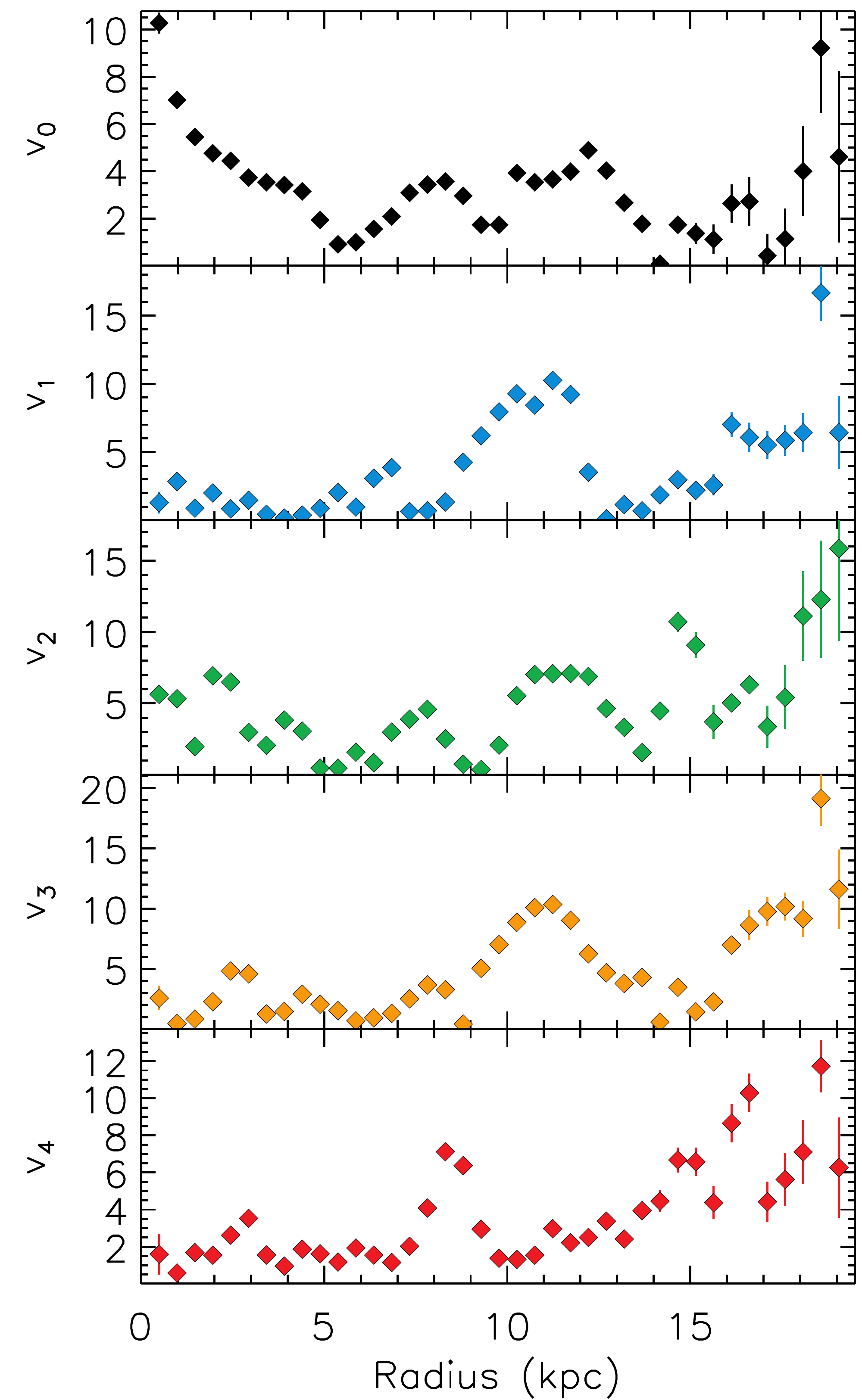}
\caption{Decomposition of the  \hi\  velocity field of M33 into Fourier coefficients.  
 The amplitudes  v$_k$ (in \kms) of each kinematical Fourier  $k$-mode 
 are shown.}
 \label{fig:fouriervfobs}
\end{centering} 
\end{figure}

We restrict the model to $k=4$, implying that  a perturbation of the potential of order of up to $m=3$, 
maybe $m=5$, is likely to be detected, if it exists. 
 For the harmonic model  the inclination and position angle were fixed at the values of the tilted-ring model of Sect.~\ref{sec:kinparam}. 
 The angular sampling is  $\sim$490 pc (120\arcsec). 
We derived the amplitudes of the noncircular and asymmetric motions, given respectively  by ${\rm v_1}=|s_1|$ and ${\rm v_k}=(c_k^2+s_k^2)^{1/2}$ for $k>1$,
 and ${\rm v_0}=\sqrt{(c_0-{\rm V_{sys}})^2}$.  These amplitudes are shown in Fig.~\ref{fig:fouriervfobs}. The uncertainties are the 
 $1\sigma$ formal errors from the fit.
 
One first observes  increasing amplitudes for every $k$ terms at $R \gtrsim 15$ kpc.    
 These may be other signatures of the interaction with M31 on the outer M33 velocity field. 
    Interestingly, a similar trend was  observed for v$_2$, v$_3$ and v$_4$ in the outer velocity field 
    of another grand-design spiral galaxy, Messier 99, which is also perturbed by its environment \citep{Chemin2016}. 

 Then,  the variations of v$_0$ and v$_2$ are similar. Larger amplitudes are observed in the inner 5 kpc, as well as at $R\sim 8$ and 12 kpc.  
 This is evidence for a kinematical lopsidedness, i.e. a $m=1$ perturbation of the M33 gravitational potential.
   Larger $k=0$ and $k=2$ terms in the innermost disk region are other similarities with M99. The particularity of M33,
    however, is that the difference  between
    the rotation curves for the approaching and receding disk sides ($R<5$ kpc) is not as prominent as in M99. \citet{Chemin2016} showed from an asymmetric 3D mass model that 
   the central  peak of these even terms cannot be modeled solely by a lopsided  gravitational potential of luminous matter, and proposed that  
    dark matter may be lopsided as well in the innermost regions of M99. We refer to a future work 
    a similar modeling of asymmetric mass distribution 
     of M33 to investigate whether lopsided gas and stellar potentials are enough to explain the kinematical asymmetry.
     
 Second, the variations of v$_1$ and v$_3$ are similar. 
The $k=1$ and $k=3$ terms are small inside $R = 7$ kpc and larger at $R \sim 11$ kpc, within the zone of strong major axis PA variation.  
Smaller values in the inner disk reflects perfectly the observation 
that no  bisymmetric structure dominates in the inner density map  (Fig.~\ref{fig:velomap}).
On another hand, larger values at large radius reflects the dynamical impact of the M33 warp.

  Thirdly, a bump is evidenced for the $k=4$ term at $R \sim$ 8 kpc, coincinding with those seen for the $k=0$ and $k=2$ terms. 
 Whether this feature is caused by the same $m=1$ perturbation is 
 something that remains unclear.  It is unlikely to be caused by higher order perturbations, as no   $m=3$ or $m=5$  
 modes are observed in the gaseous and stellar density maps. 

 The average amplitude of the noncircular motion is  $\langle {\rm v}_1 \rangle = 3.8 \pm 0.6$ \kms.
 The average amplitude of asymmetries are $\langle {\rm v}_0 \rangle = 3.3\pm 0.3$ \kms, $\langle {\rm v}_2 \rangle = 4.9 \pm 0.6$ \kms, 
 $\langle {\rm v}_3 \rangle =4.9 \pm 0.7$ \kms, and $\langle {\rm v}_4 \rangle =3.7 \pm 0.4$ \kms, again showing 
the more important impact of $m=1$ and $m=2$ perturbations on the gravitational potential of M33. The level of such asymmetries is 
consistent with the asymmetry measured between the rotation curves of the approaching and receding disk sides.

\subsection{Comparisons with previous works}
\label{sec:comprc}

\subsubsection{DRAO-Arecibo versus VLA-GBT}

 \begin{figure}[t] 
\centering
 \includegraphics[width=\columnwidth]{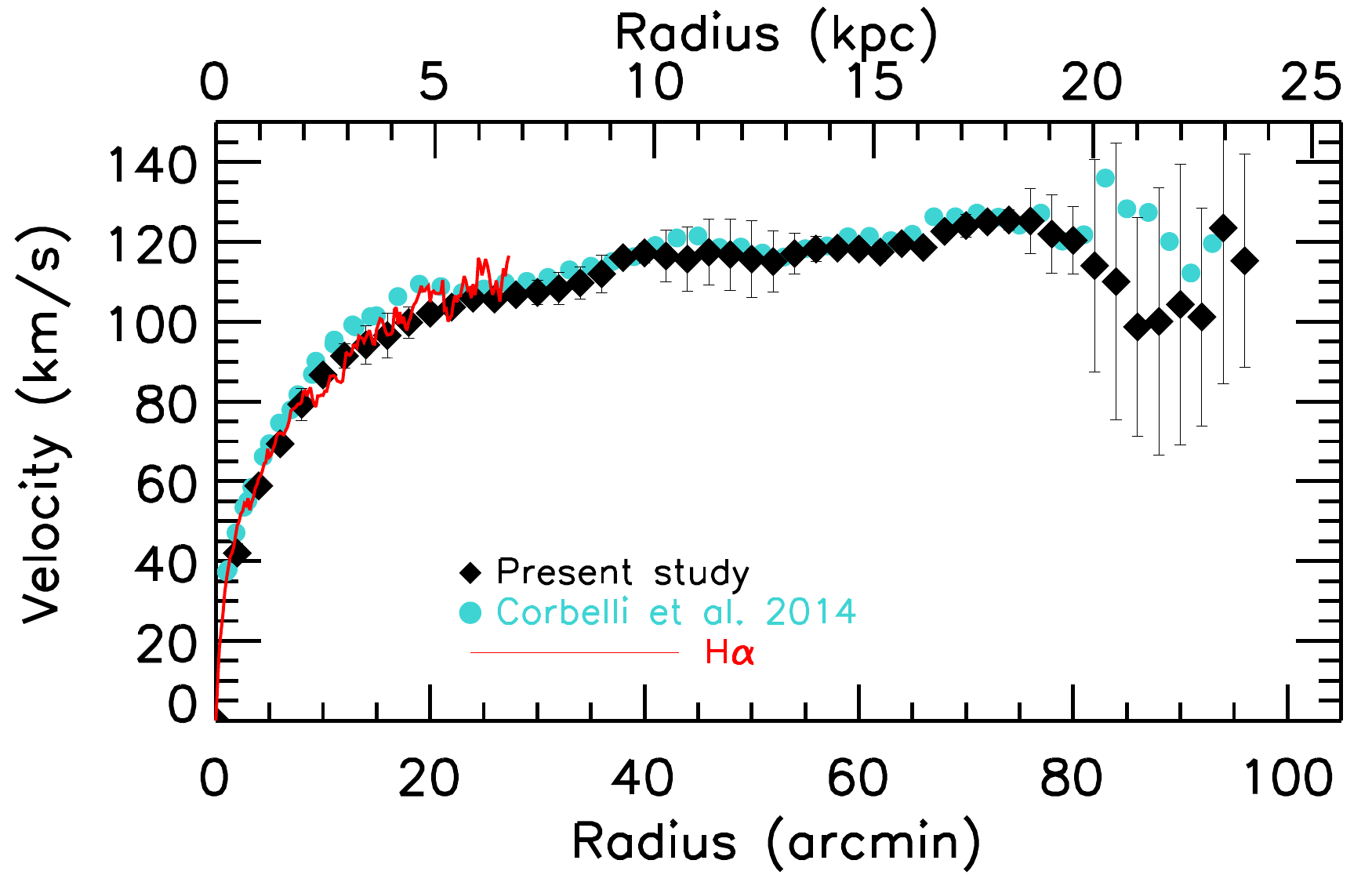}
\caption{\label{fig:rv_comp} Comparison bewteen our new \hi\ rotation curve 
(black diamonds) with the \ha\ rotation curve 
from Kam15 (red line) and the \hi\ rotation curve from \citet{Corbelli2014} (cyan 
circles).}  
\end{figure}

The significant twist of the  major axis of the \hi\ velocity field we measure at large radius from our DRAO-Arecibo dataset 
is very consistent with results obtained from VLA-GBT data by \citet{Corbelli2014}, or from older Arecibo 21-cm measurements
 by \citet{Corbelli1997}. 
However the comparison with the kinematical tilt found in \citet{Corbelli2014}
 is made difficult because these authors found two different trends of 
 inclination variation at large radii. On the one hand \citet{Corbelli2014} found  a disk inclination that decreases, 
 as based on a three-dimensional modeling of the \hi\ datacube. 
 On the other hand they found a slightly increasing inclination from more 
 traditional tilted-ring models of the velocity field. 
 Our warp model is therefore in agreement with this part of their modeling, but not entirely with their 3D datacube modeling.

Comparing our rotation curve with the one from \citet{Corbelli2014}  
is not straightforward as well because their curve was derived from these two different models of tilted disk. 
The DRAO and VLA rotation curves are shown in Fig.~\ref{fig:rv_comp} and agree well within the uncertainties for $R \le 20$ kpc.  
Then the shapes unsurprinsingly differ beyond  $R = 20$ kpc. 
A more detailed investigation shows that when we compare with the curve they obtained from a \textsc{rotcur}-like model similar to ours, 
the shapes become comparable at these radii, showing a drop out to $R \sim$22 kpc, followed by a rise. 
Therefore it is the action of the 3D modeling  they made of the datacube that causes the shape difference at the largest radii.

What is clear, however, is that both  \hi\ rotation curves  
are perturbed at these radii, harbouring larger scatter and uncertainties. 
We conclude that the \hi\ rotation curve of M33 is only reliable out to $\sim$20 kpc, 
irrespective of the method used to derive it. Fortunately,
this outermost region has negligible impact on the mass distribution 
models described in Sect.\ref{sec:massmod} because we used as weighting function 
of the rotation velocities the inverse of the squared errors.

\subsubsection{The hybrid \ha\ - HI rotation curve of M33}
\label{sec:adopRC}
  
Figure~\ref{fig:rv_comp} also shows the very good agreement between the DRAO \hi\ rotation curve with that of the ionized gas disk from Kam15. 
The \ha\ rotation curve naturally harbours more wiggles than the 21-cm data because of its higher resolution.

The analysis of the mass distribution (Sect.~\ref{sec:massmod}) 
is based on a hybrid rotation curve which is obtained by combining the \ha\ velocities from Kam15 for $R \le 6.5$ kpc 
 with the \hi\ velocities for $R>6.5$ kpc.  This radius was chosen 
 sligthly before the location from where the \hi\ warping starts.
The advantages of using a hybrid curve for mass models are to benefit from the high sampling 
 of the \ha\ data (20 pc, 5\arcsec) to constrain the central velocity gradient more accurately 
than with our radio interferometry only (490 pc, 120\arcsec\ sampling), and from the large extent of the neutral gas disk.  
The \ha\ rotation curve within $R = 6.5$ kpc has $\sim 10$ times more velocity points than the \hi\ curve 
for $R>6.5$ kpc, yielding a final rotation curve with 353 data points at a hybrid resolution of 20 pc-490 pc. 
The hybrid curve is listed in Tab.~\ref{tab:hybridrc} of   
Appendix~\ref{sec:hybridrc}.

%% file: Massmodel.tex
The modeling of the mass distribution fits a model velocity profile to the hybrid \ha-\hi\ rotation curve of Messier 33. 
The contributions from the luminous matter are those of the 
gaseous and stellar disks, inferred from mass 
surface density profiles (\S\ref{sec:stellar} and \S\ref{sec:atommolec}). 
 We have considered models with dark matter (DM) using two different  forms for the 
DM halo:  the pseudo-isothermal sphere (ISO) and the Navarro-Frenk-White (NFW) models (\S\ref{sec:massmod_DM}). 
We also did the modeling within the framework of Modified Newtonian Dynamics (\S\ref{sec:massmod_MOND}). 

\subsection{Stellar component}
	\label{sec:stellar}

In Kam15, the M33 stellar bulge was shown to have a negligible impact in the mass distribution \citep[see also][]{Corbelli2007,Corbelli2014}.
 Therefore only the velocity contribution from the stellar disk, $\rm V_\star$, is needed in the present study. 
It has been derived with the task \textsc{rotmod} of \textsc{gipsy}  \citep{vanderHulst1992} 
from near-infrared surface photometry, which is well-known to give the best representation of the old stellar disk population that  
 contributes the most to the stellar mass.  
Kam15 derived the surface brightness profile of M33 at 3.6 $\mu$m from Spitzer/IRAC data, where contaminating bright stars have been removed.  
More complete details on the derivation of the surface brightness profile are given in Kam15. 
The disk mass-to-light ratio at 3.6 $\mu$m, $\Upsilon$, has been estimated from infrared colours and stellar population synthesis
(SPS) models 
following the prescriptions given by \cite{Oh2008} and \cite{deBlok2008}:
\begin{equation}
\Sigma [ {\rm M_{\odot}\ pc^{-2}}] = \Upsilon \times 10^{-0.4\, (\mu_{3.6}-C_{3.6})} 
\label{eq:massdensity} 
\end{equation}
where $\mu_{3.6}$ is the surface brightness
 and  $C_{3.6}=24.8$ is a correction value, as given in \cite{Oh2008}.  
Using the $J-K$ colour index from \cite{Jarrett2003}, Kam15 derived  $\Upsilon = 0.72 \pm 0.1$. 
This corresponds to a stellar mass of $(7.6 \pm 1.1)\ 10^9$
\msol, which is 38\% (58\%, respectively)
 higher than the fixed (best-fit) stellar mass  
 given in \citet{Corbelli2014}. Note that $\Upsilon \sim 0.7$ 
 is the upper limit expected from mass distribution modeling of 
 the large galaxy sample of \citet{Lelli2016}.

 This discrepancy is the reason why we have also performed  models with $\Upsilon=0.52$ 
to facilitate comparisons with the  results of \citet{Corbelli2014}. 
This value of $\Upsilon$ was obtained by scaling the maximum of
 the stellar velocity curve 
to $\sim 70$ \kms, which corresponds to the highest contribution from stars in the fixed stellar mass fits of \citet{Corbelli2014}. 
 $\Upsilon=0.52$ is valid here
  in the Spitzer/IRAC 3.6 $\mu$m  band only, and corresponds to a stellar mass of 
  $5.5\ 10^9$ \msol, which thus agrees perfectly with the value from the SPS model of \citet{Corbelli2014}. 
  Assuming $\Upsilon=0.52$ that is constant with radius in our models  
 is a good way to account for a stellar velocity contribution fairly similar in every aspects to \citet{Corbelli2014}. 
 Note also that it corresponds to the mean value found for the sample of \citet{Lelli2016}, and appears more in agreement with 
 most of galaxies of their sample as bright as M33 than $\Upsilon=0.72$.

\subsection{Atomic and molecular gas components}
\label{sec:atommolec}
The velocity contribution from the disk of neutral gas, $\rm V_{atom}$, is derived from  our deep DRAO+Arecibo dataset 
by integrating the 
 total atomic gas mass surface densities (Fig.~\ref{fig:radprof}) multiplied by a factor 
of $\sim$1.3 to take into account the contribution from Helium. 

The molecular gas, traced by the CO line, is mainly concentrated in the innermost kpcs.  Molecular gas in M33 has been observed 
by \cite{Tosaki2011} with  the Nobeyama Radio Observatory at 19\arcsec\ resolution, and by \cite{Gratier2010} and \cite{Druard2014} with the IRAM 30-m dish at a resolution of
 $\sim 12\arcsec -15\arcsec$.  \cite{Tosaki2011} report that the \hI\ and CO peaks are not always correlated and  that the density of the 
atomic gas is higher than that of the molecular gas in the inner parts, while   \cite{Gratier2010}  showed that 
the \hi\ density is lower than that of the molecular gas. 
 Using a conversion factor of $\rm N(H_2)/I_{CO(1\rightarrow 0)} = 4\times10^{20} cm^{-2}/$(K~\kms), 
 twice the  value found for the Milky Way (M33 is half the solar metallicity),  they measured
 an average   density of $\Sigma_{H_2} =8.5 \pm 0.2$~\msol\ $\rm pc^{-2}$ for the central kpc, and a total molecular gas mass of $\sim 3.3\times10^8$ \msol\ for the entire M33 disk. 
   We used the \hdeux\ density profile derived from  \cite{Druard2014}, scaled  by a factor of $\sim 1.3$ to infer
    the velocity contribution of the total molecular gas component, $\rm V_{mol}$. 
    
\subsection{Dark Matter halo component}
\label{sec:massmod_DM}
 
The  total  rotation velocity $\rm V_{rot}$ in the models with a DM component is defined by: 
\begin{equation}
\label{eq:vtot}
	\rm  V_{rot}^2 =   V_\star^2 + V_{gas}^2 + V_{DM}^2  
\end{equation} 
where $\rm V_\star$ is the contribution from the  stellar disk and 
$\rm V_{\rm gas}=(V_{atom}^2 + V_{mol}^2)^{1/2}$ the total contribution from the gaseous disk, as deduced in 
\S\ref{sec:atommolec}, and $\rm V_{DM}$  from a DM halo assumed spherical.
 
\subsubsection{The pseudo-isothermal sphere model} 
\label{sec:DM_iso} 

Here, the density profile of  DM  is given by:
\begin{equation}
\rho_{\rm ISO}(R) = \frac{\rho_{0}}{1 + (R/R_c)^{2}}
\end{equation}
 The corresponding circular velocities are:
\begin{equation}
{\rm V_{ISO}}(R) = \sqrt{4\pi G \rho_0 R_c^2 \left(1 - R/R_c\arctan(R/R_c)\right)}
\end{equation}
where $\rho_{0}$ and $R_c$ are the central density and the core radius of the halo, respectively.
We can describe the steepness of the inner mass density profile by a power law $\rho \sim R^\alpha$.
In the case of the ISO halo,  $\alpha \rightarrow 0$ (the halo core has a density almost constant).

\subsubsection{The Navarro-Frenk-White model}
\label{sec:DM_nfw} 

The NFW model -- the so-called ``universal halo'' -- is deduced from Cold Dark Matter simulations \citep{Navarro1997}.
 The density profile is cuspy, following a $\rho \propto R^{-1}$ law in the center, and is given
  by:  

\begin{equation}
\rho_{\rm NFW}(R)={\rho_i \over R/R_s (1+R/R_s)^2}
\label{eq:ro_nfw}
\end{equation}
where $\rho_i \thickapprox 3 {\rm H_0^2}/(8 \pi G)$ is the critical density for closure of
the Universe and $R_s$ is a scale radius. The velocity contribution corresponding to
this halo  is given by:  
\begin{equation}
{\rm V_{NFW}}(R)=\rm V_{200}~\sqrt{{\ln(1+cx)-cx/(1+cx)} \over {x(\ln(1+c) -c/(1+c))}}
\label{eq:v_nfw}
\end{equation}
 with $\rm V_{\rm 200}$ that is  the velocity at a radius $R_{\rm 200}$ at which the density is 200 times that for closure of the Universe,   
  $c = R_{\rm 200}/R_s$ gives the concentration parameter of the halo and $x = R/R_s$.

 \subsection{Modified Newtonian Dynamics mass models}
 \label{sec:massmod_MOND}
 
An alternative to dark matter to explain the missing mass problem is MOdified Newtonian Dynamics \citep{Milgrom1983b,Milgrom1983a}. 
MOND has been successful to reproduce correctly many galaxy rotation curves \citep[e.g.][]{Sanders1998,Gentile2010}. 
It postulates that in a regime of acceleration much smaller than a universal constant acceleration, $a_0$, the classical Newtonian dynamics is no more valid 
and the law of gravity is modified. 

In the MOND framework, the gravitational acceleration of a test particle is given by :
\begin{equation}
 \mu(x = g/a_{0}) g = g_{N}
\end{equation}
where $g$ is the  acceleration,  $g_{N}$ the Newtonian acceleration, and  $\mu(x)$ is an interpolating function that must satisfy: 
$\mu(x) = x$ for $x << 1$ and $\mu(x) = 1 $ for $ x >> 1$. The MOND velocity profile thus depends on $\mu(x)$. 
For the ``standard'' $\mu$-function proposed by \cite{Milgrom1983b}   
\begin{equation}
\mu(x)= {x\over \sqrt{1+x^2}}
\label{eq:mondstandard}
\end{equation}
the MOND rotation velocity is:

\begin{equation}
\rm V_{STD} =  V_{lum} \sqrt{ \sqrt{0.5\ \left(1+ \sqrt{ 1 + (2a_0 R/{\rm V^2_{lum}})^2} \right)} }
\label{eq:v_mondstd}
\end{equation}
with $\rm V_{lum}^2= V_\star^2 + V_{gas}^2$,   $\rm V_\star$ and $\rm V_{gas}$ being as in Eq.~\ref{eq:vtot}.
The standard scale acceleration is $a_0=  \rm (H_0/75)^2 \times 1.2\ 10^{-8} = 0.99$ cm s$^{-2}$ ($\rm H_0=68$ \kms\ Mpc$^{-1}$). 

For the simple function 
\begin{equation}
\mu(x)= {x\over 1+x}
\label{eq:mondsimple}
\end{equation}
that was shown to apply better to galaxy rotation curves \citep{Famaey2005}, the velocity is given by
\begin{equation}
\rm V_{SIM} = V_{lum} \sqrt{1+0.5\left(\sqrt{1+4a_0R/{\rm V^2_{lum}}}-1\right)}
\label{eq:v_mondspl}
\end{equation}

Both $\rm V_{STD}$ and $\rm V_{SIM}$ models were fitted to the rotation of M33, with free $a_0$ 
and fixed or free  $\Upsilon$. 

  \begin{figure*}[th]
 \begin{centering}
  \includegraphics[width=8cm]{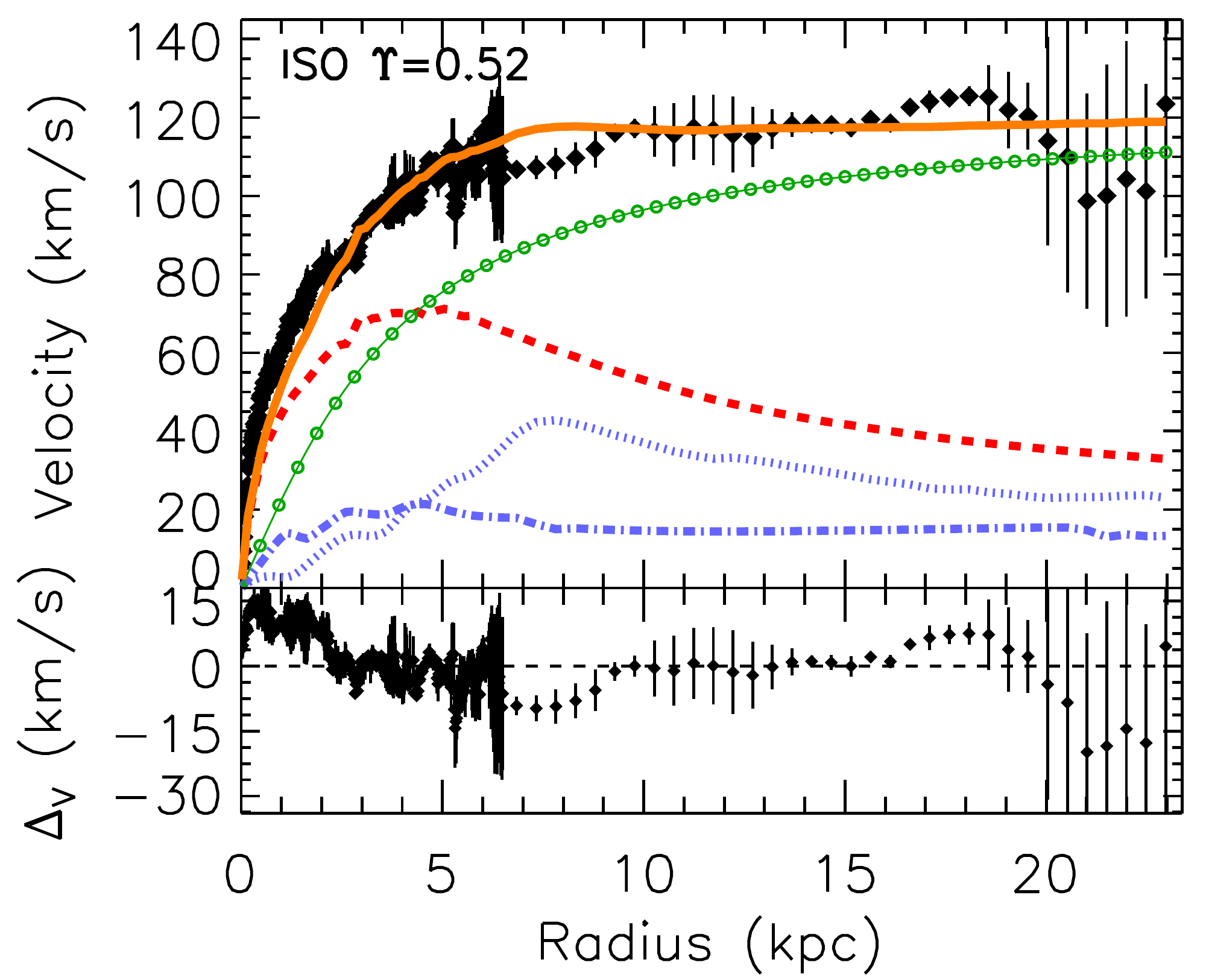}\includegraphics[width=8cm]{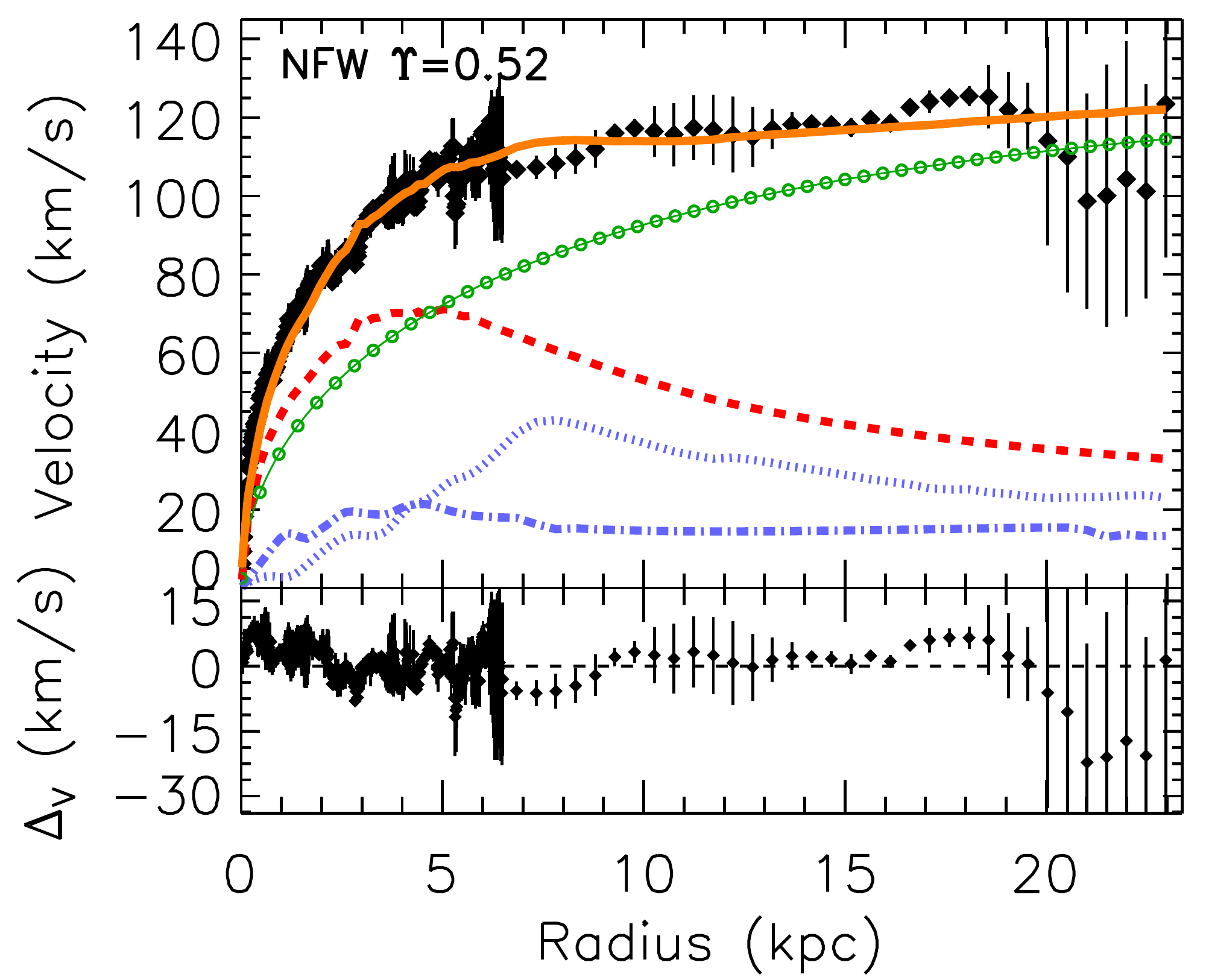}
  \includegraphics[width=8cm]{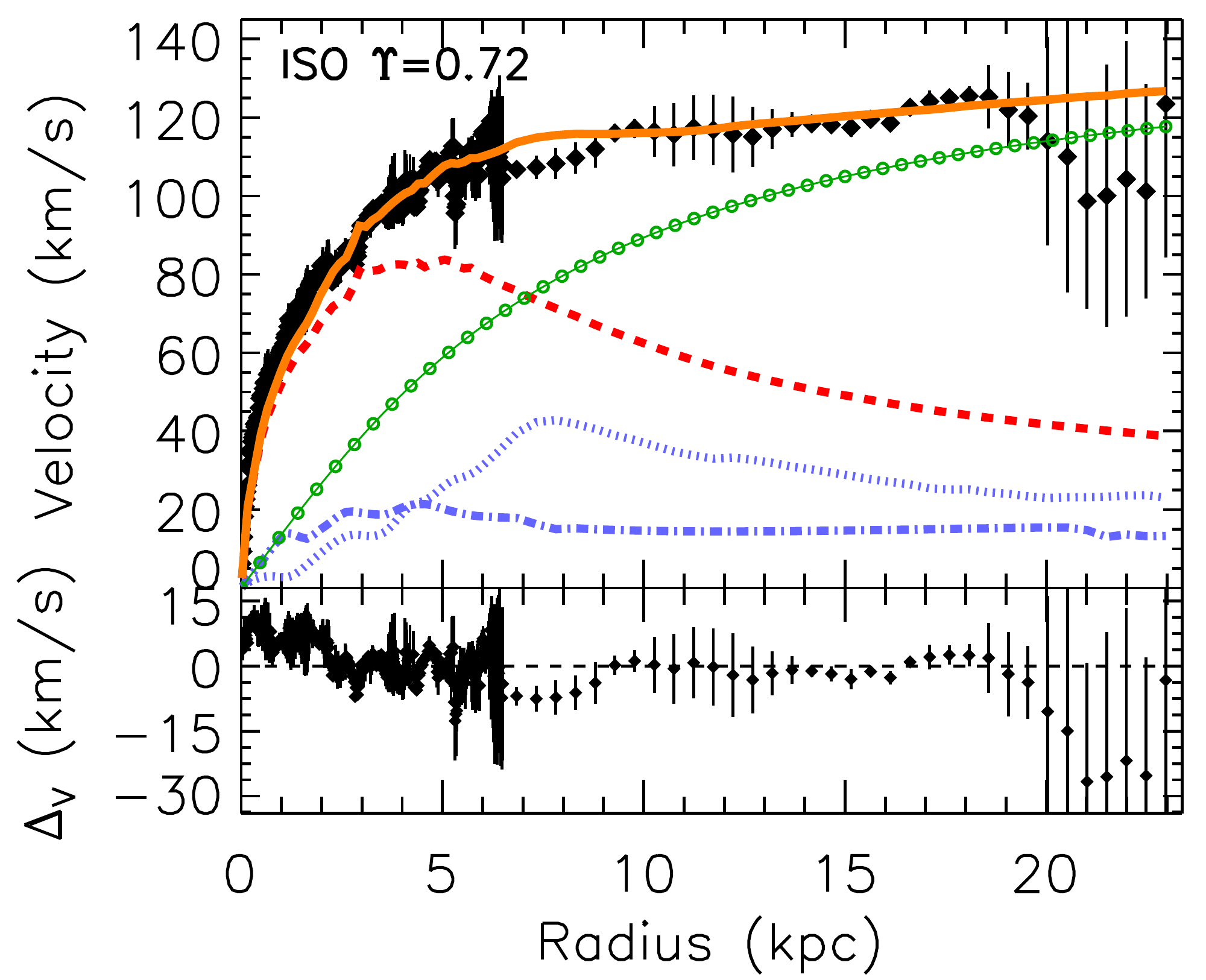}\includegraphics[width=8cm]{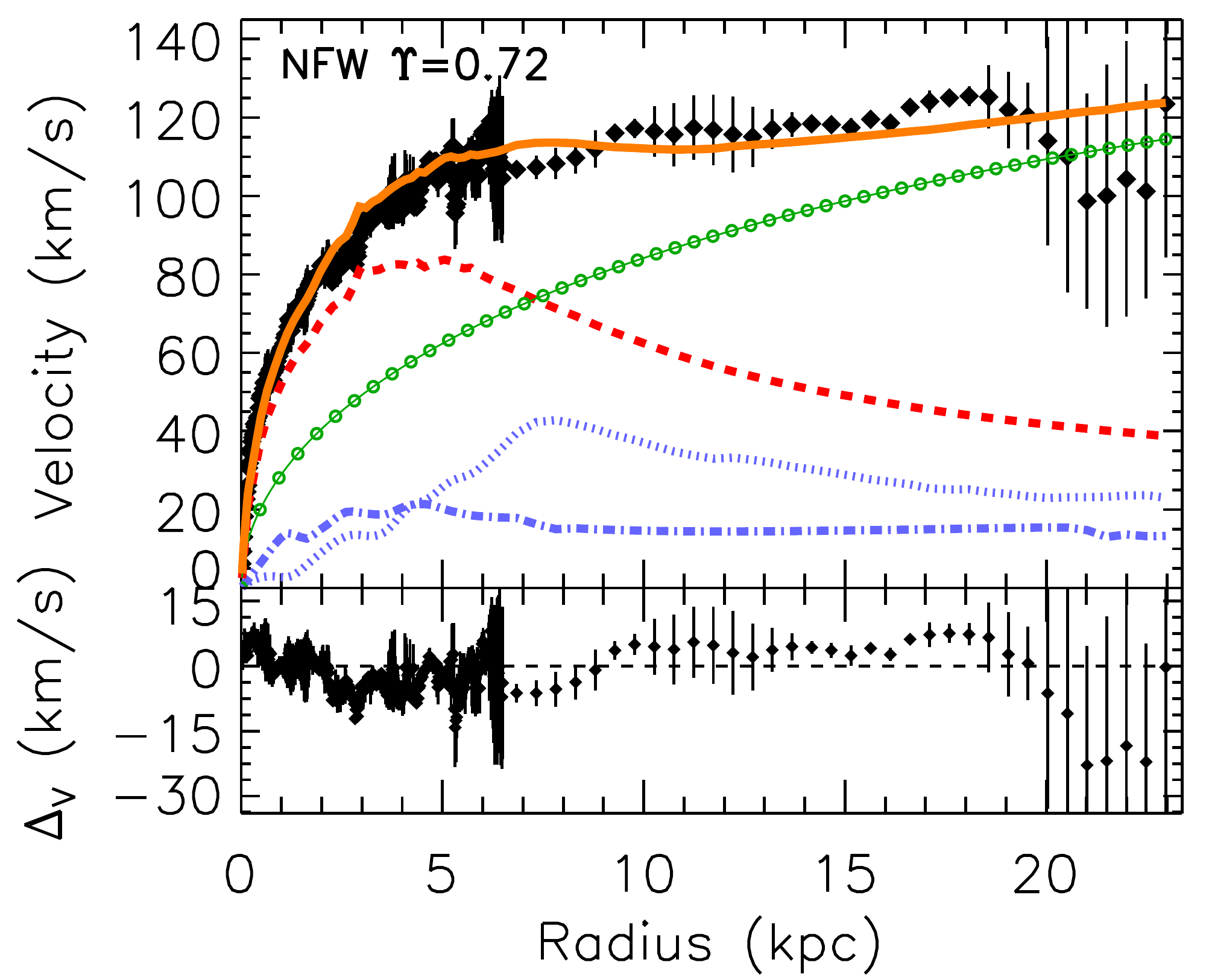}
  \includegraphics[width=8cm]{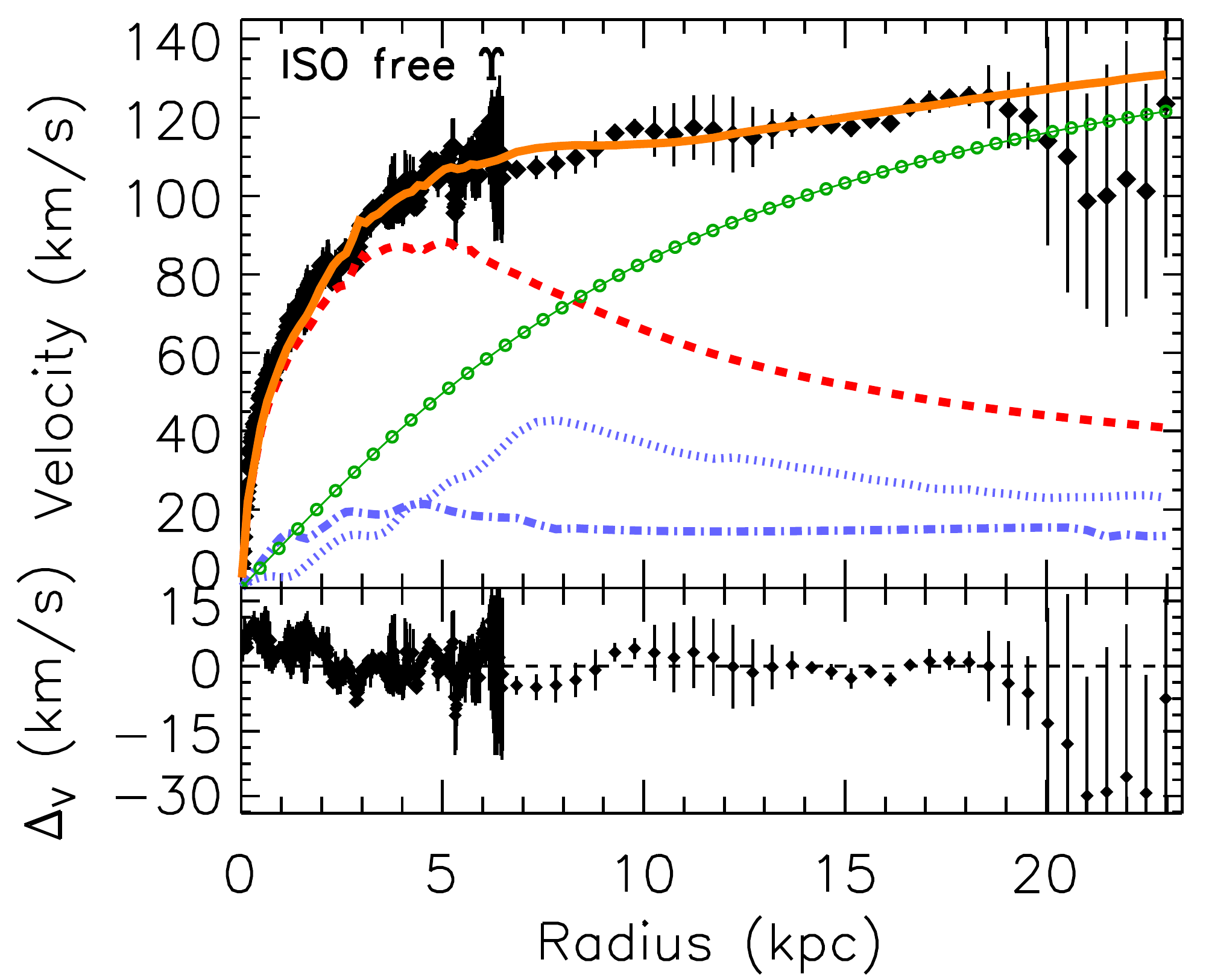}\includegraphics[width=8cm]{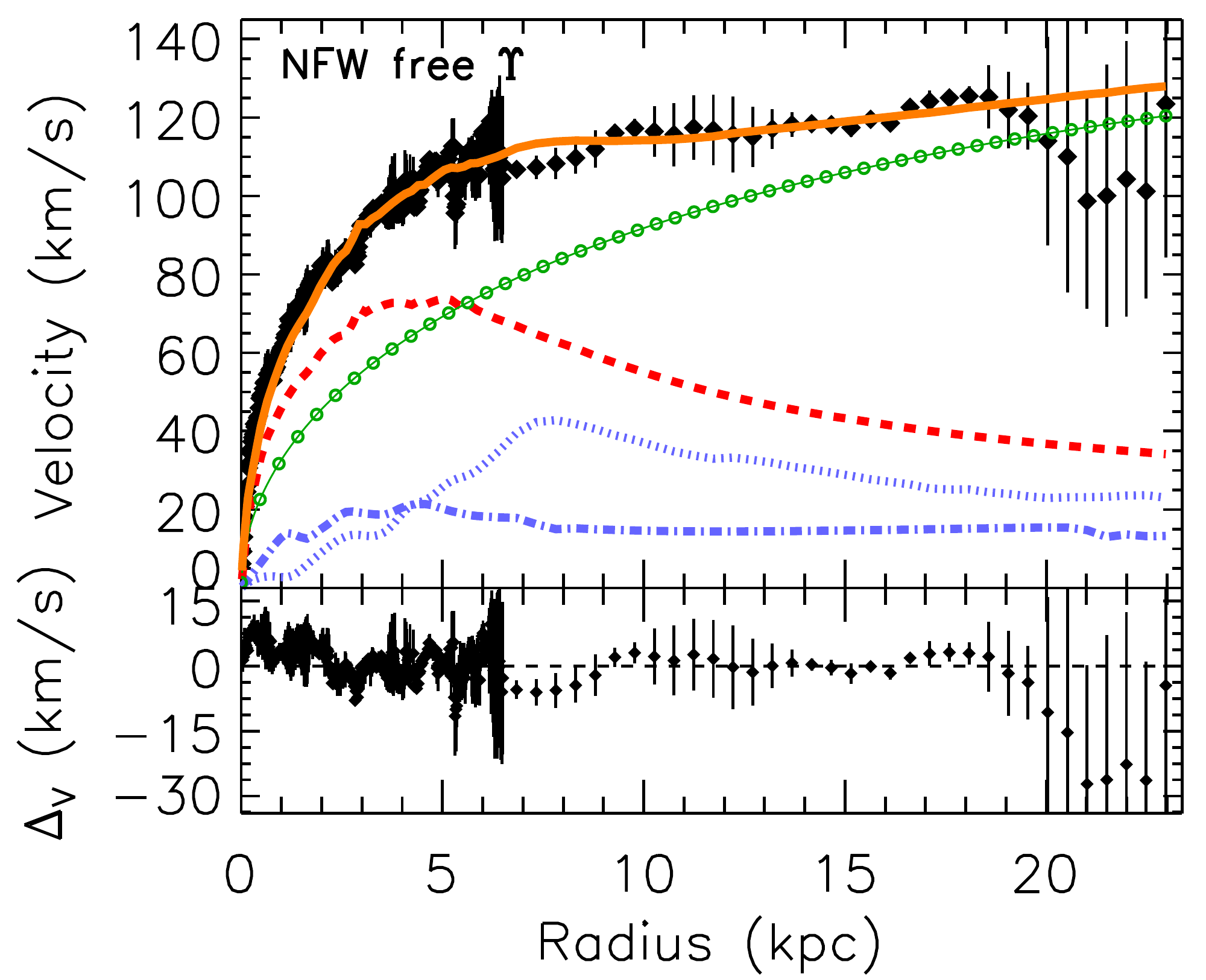}
 \caption{Mass distribution models of M33 with the ISO (left column) and NFW (right column) haloes. 
 From top to bottom, results are shown for different values of the stellar disk mass-to-light ratio: 
 fixed $\Upsilon=0.52$, fixed $\Upsilon=0.72$ and free, best-fit $\Upsilon$, where the fixed values 
 were inferred from stellar population models (see text).
  Black filled symbols represent the observed data, a solid orange line the model of the total velocity curve, a dashed 
  red line the contribution from the stellar disk, dotted and dashed-dotted blue lines those from the atomic and molecular gas 
  disks, respectively, and a circle green line that from the dark matter halo.  For each sub-panel, the bottom insert shows the velocity 
  residual velocity curve $\Delta_{\rm V}$ (observed minus modeled rotation curves).}
  \end{centering}
  \label{fig:massmoddm}
 \end{figure*}

  \begin{figure*}[th]
 \begin{centering}
\includegraphics[width=0.33\textwidth]{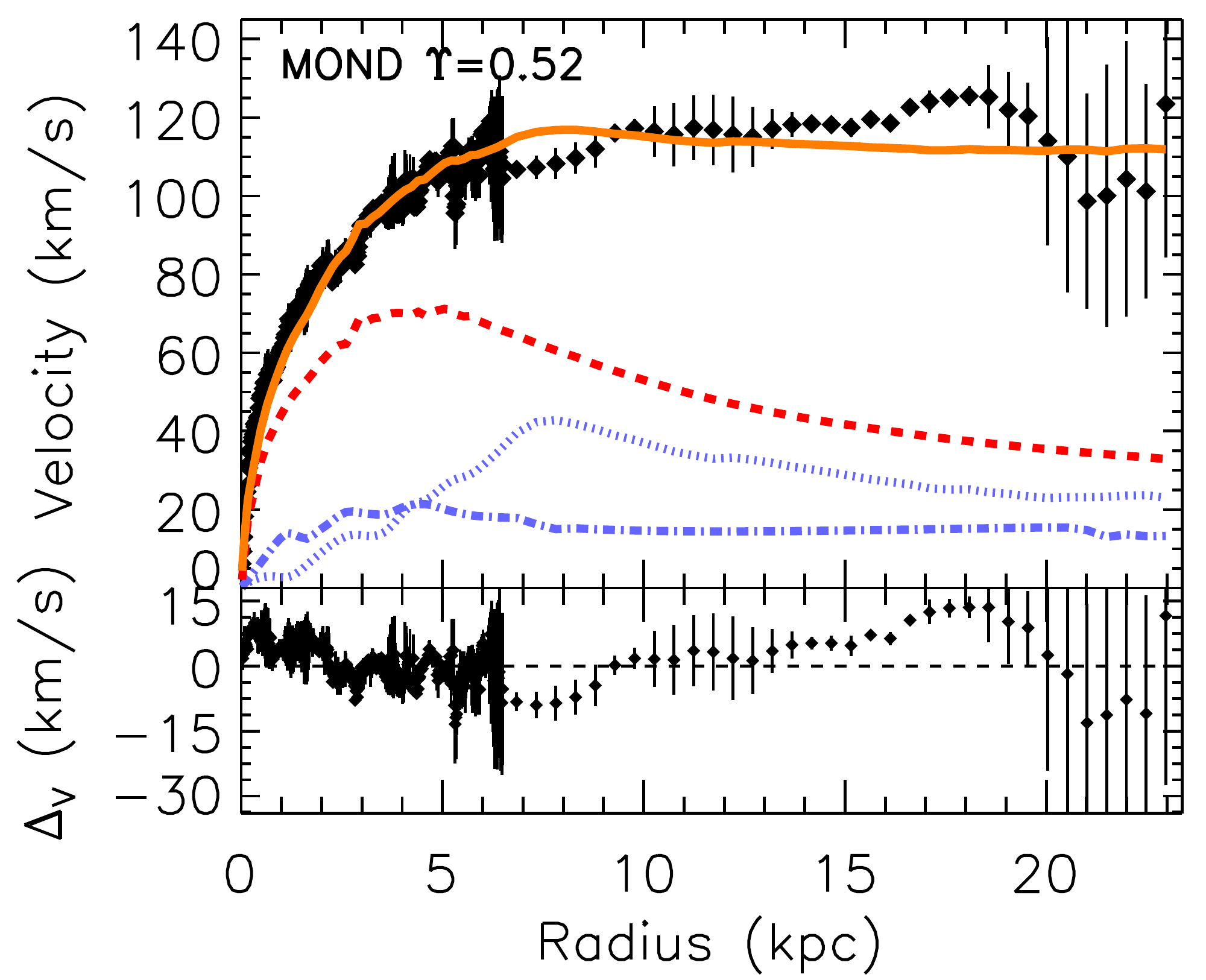}\includegraphics[width=0.33\textwidth]{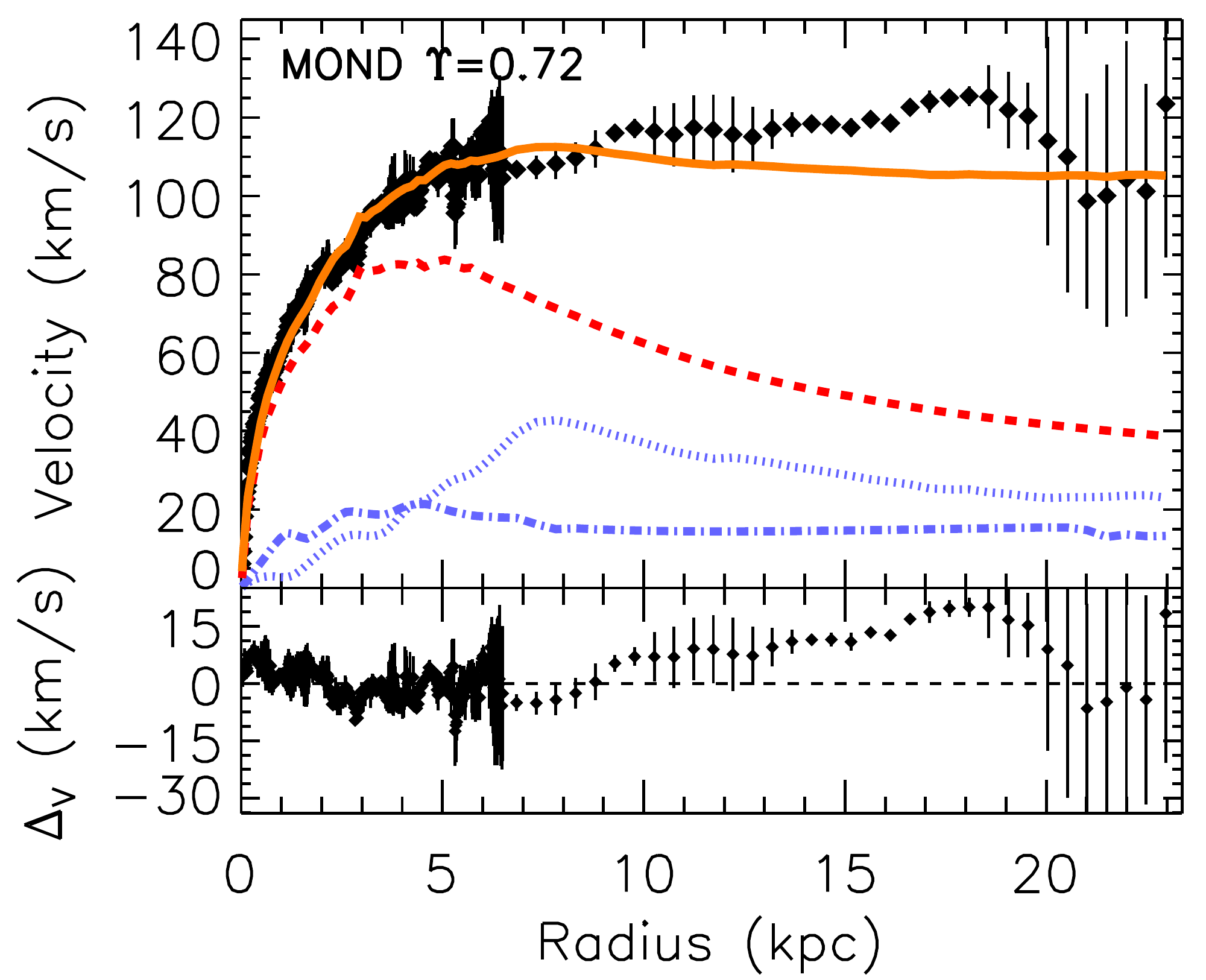}\includegraphics[width=0.33\textwidth]{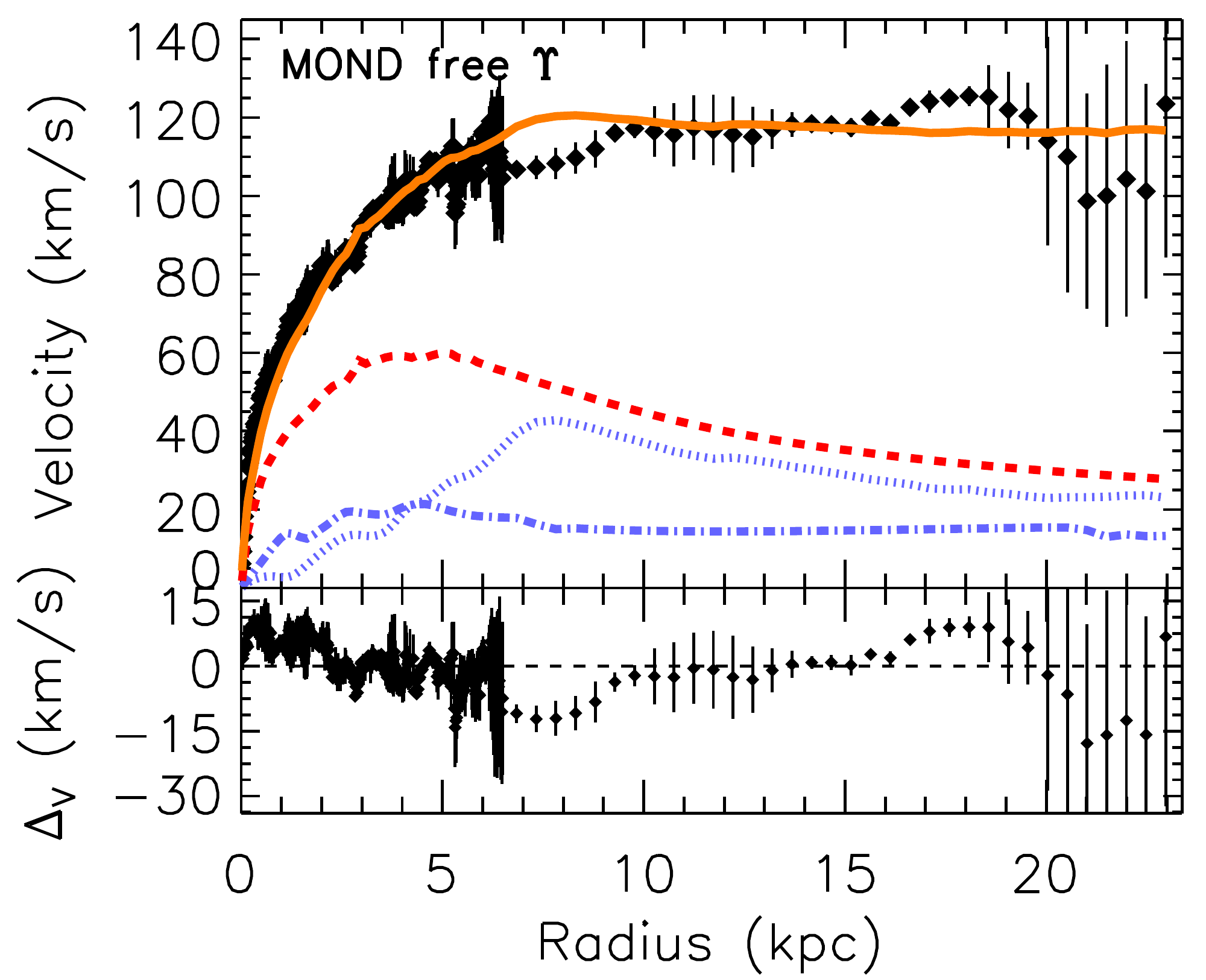}
 \caption{Mass distribution models of M33 with MOND (standard interpolation function). Symbols and lines are the same as in Fig.~\ref{fig:massmoddm}.}
 \label{fig:massmodnodm}
  \end{centering}
 \end{figure*}

\begin{table}
\begin{center}
\caption{Results of the mass models.}
\label{tab:resmm}
\begin{tabular}{l|c|c|c|c}
\hline\hline 
Model  &  Parameter  & $\Upsilon=0.52$& $\Upsilon=0.72$   &  Free $\Upsilon$     \\ 
\hline 
ISO             & $\rho_0$		  & $29.8 \pm 1.1$	& $10.6 \pm 0.3$      &  $6.5 \pm 0.6$      \\ 
        	& $R_c$ 		  & $3.1 \pm 0.1$	& $6.1 \pm 0.2$       &  $8.9 \pm 0.7$      \\		       
        	& $\Upsilon$  	  & 0.52		 & 0.72 	        &  $0.80 \pm 0.01$   	\\   
\hline
NFW 	        &$V_{200}$	       	  & $114.4 \pm 0.7$	& $152.5 \pm 0.4$     &  $139.9 \pm 0.5 $    \\
		& $c$ 			  & $6.05\pm 0.05$	& $3.44 \pm 0.01$     &  $ 4.55 \pm 0.01 $   \\
		& $\Upsilon$	  &	 0.52		 & 0.72 	        &  $0.56 \pm 0.01 $  	  \\		      
\hline
MOND  	        & $a_0$	          & $1.22\pm 0.01$    	& $0.78\pm 0.01$      & $1.76 \pm 0.11$ \\
Standard        & $\Upsilon$      & 0.52                 &  0.72         	& $0.50\pm 0.03$ \\
\hline
MOND  	        & $a_0$	          & $0.68\pm 0.01$    	& $0.33\pm 0.01$      & $1.84 \pm 0.11$ \\
Simple          & $\Upsilon$      & 0.52                 &  0.72         	& $0.35\pm 0.02$ \\
\hline 
\end{tabular} 
\end{center}
Comments:
 $\Upsilon$ is in \msol/\lsol,  $\rho_0$ and $\rho_{-2}$ in $10^{-3}$ \msol\ pc$^{-3}$,   
$R_c$ and $R_{-2}$  in kpc, $V_{200}$ in \kms. 
For the MOND model, $a_0$ is in units of $10^{-8}$ cm s$^{-2}$. 
Uncertainties are statistical (formal) 1$\sigma$ errors from the fits.
\end{table}

\begin{table*}
\begin{center}
\caption{Comparison between mass models.}
\label{tab:deltachi2}
\begin{tabular}{l|r|l}
\hline\hline 
\multicolumn{1}{c|}{Modeling with Dark Matter} & \multicolumn{1}{c|}{Value}  & \multicolumn{1}{c}{Diagnosis}\\
\hline
$ (\chi_{\rm ISO,\Upsilon=0.52}^2-\chi_{\rm NFW,\Upsilon=0.52}^2)/\chi_{\rm NFW,\Upsilon=0.52}^2 $ & 0.52  &  NFW,$\Upsilon=0.52$ more likely than ISO,$\Upsilon=0.52$   \\
$ (\chi_{\rm ISO,\Upsilon=0.72}^2-\chi_{\rm NFW,\Upsilon=0.72}^2)/\chi_{\rm NFW,\Upsilon=0.72}^2 $ & $-0.68$      &  ISO,$\Upsilon=0.72$ more likely than NFW,$\Upsilon=0.72$   \\
$ (\chi_{\rm ISO,free \Upsilon}^2-\chi_{\rm NFW,free \Upsilon}^2)/\chi_{\rm NFW,free \Upsilon}^2 $ & 0.13  &	 NFW, free $\Upsilon$ more likely than ISO, free $\Upsilon$   \\
\hline
$ (\chi_{\rm NFW,\Upsilon=0.72}^2-\chi_{\rm NFW,\Upsilon=0.52}^2)/\chi_{\rm NFW,\Upsilon=0.52}^2 $ & 0.52   & NFW,$\Upsilon=0.52$ more likely than NFW,$\Upsilon=0.72$    \\
$ (\chi_{\rm ISO,\Upsilon=0.72}^2-\chi_{\rm ISO,\Upsilon=0.52}^2)/\chi_{\rm ISO,\Upsilon=0.52}^2 $ & $-0.67$      &  ISO,$\Upsilon=0.72$ more likely than ISO,$\Upsilon=0.52$   \\
\hline\hline
\multicolumn{1}{c|}{Modeling with MOND} & \multicolumn{1}{c|}{Value}  & \multicolumn{1}{c}{Diagnosis}\\
\hline
$ (\chi_{\rm SIM,\Upsilon=0.52}^2-\chi_{\rm STD,\Upsilon=0.52}^2)/\chi_{\rm STD,\Upsilon=0.52}^2 $ & 0.43  &  STD,$\Upsilon=0.52$ more likely than SIM,$\Upsilon=0.52$ 	  \\
$ (\chi_{\rm SIM,\Upsilon=0.72}^2-\chi_{\rm STD,\Upsilon=0.72}^2)/\chi_{\rm STD,\Upsilon=0.72}^2 $ & 0.52  &  STD,$\Upsilon=0.72$ more likely than SIM,$\Upsilon=0.72$ 	 \\
$ (\chi_{\rm SIM,free \Upsilon}^2-\chi_{\rm STD,free \Upsilon}^2)/\chi_{\rm STD,free \Upsilon}^2 $ & 0.11   & STD, free $\Upsilon$ more likely than SIM, free $\Upsilon$ \\
\hline
$ (\chi_{\rm STD,\Upsilon=0.72}^2-\chi_{\rm STD,\Upsilon=0.52}^2)/\chi_{\rm STD,\Upsilon=0.52}^2 $ & 0.33  &  STD,$\Upsilon=0.52$ more likely than STD,$\Upsilon=0.72$ \\
$ (\chi_{\rm SIM,\Upsilon=0.72}^2-\chi_{\rm SIM,\Upsilon=0.52}^2)/\chi_{\rm SIM,\Upsilon=0.72}^2 $ & 0.44  &  SIM,$\Upsilon=0.52$ more likely than SIM,$\Upsilon=0.72$  \\
\hline
\end{tabular} 
~\\ Comment: STD and SIM are for MOND models with Standard and Simple (respectively) interpolation functions
\end{center}
\end{table*}   	

\begin{table*}
\begin{center}
\caption{Comparison between Dark Matter and MOND mass models.}
\label{tab:monvsdm}
\begin{tabular}{l|r|l}
\hline\hline 
 \multicolumn{1}{c|}{MOND vs DM} & \multicolumn{1}{c|}{Value}  & \multicolumn{1}{c}{Diagnosis}\\
\hline
$\rm AIC_{\rm STD,\Upsilon=0.52}-AIC_{\rm NFW,\Upsilon=0.52}$ & 7.3  &  NFW,$\Upsilon=0.52$ more likely than STD,$\Upsilon=0.52$   \\
$\rm AIC_{\rm STD,\Upsilon=0.72}-AIC_{\rm NFW,\Upsilon=0.72}$ & $-2.2$ &  STD,$\Upsilon=0.72$ more likely than NFW,$\Upsilon=0.72$   \\
$\rm AIC_{\rm STD,free \Upsilon}-AIC_{\rm NFW,free \Upsilon}$ & 5.1  & NFW, free $\Upsilon$ more likely than STD, free $\Upsilon$   \\
\hline
$\rm AIC_{\rm STD,\Upsilon=0.52}-AIC_{\rm ISO,\Upsilon=0.52}$ & $-16.5$  &  STD,$\Upsilon=0.52$ more likely than ISO,$\Upsilon=0.52$   \\
$\rm AIC_{\rm STD,\Upsilon=0.72}-AIC_{\rm ISO,\Upsilon=0.72}$ & 16.5	  &  ISO,$\Upsilon=0.72$ more likely than STD,$\Upsilon=0.72$	\\
$\rm AIC_{\rm STD,free \Upsilon}-AIC_{\rm ISO,free \Upsilon}$ & 1.9  &     ISO, free $\Upsilon$ more likely than STD, free $\Upsilon$	\\
\hline
$\rm AIC_{\rm STD,\Upsilon=0.52}-AIC_{\rm NFW,\Upsilon=0.72}$ & $-16.9$  &  STD,$\Upsilon=0.52$ more likely than NFW,$\Upsilon=0.72$   \\
$\rm AIC_{\rm STD,\Upsilon=0.72}-AIC_{\rm NFW,\Upsilon=0.52}$ & 21.9	  &  NFW,$\Upsilon=0.52$ more likely than STD,$\Upsilon=0.72$	\\
\hline
$\rm AIC_{\rm STD,\Upsilon=0.52}-AIC_{\rm ISO,\Upsilon=0.72}$ & 1.9  &  ISO,$\Upsilon=0.72$ more likely than STD,$\Upsilon=0.52$   \\
$\rm AIC_{\rm STD,\Upsilon=0.72}-AIC_{\rm ISO,\Upsilon=0.52}$ & $-1.9$  &  STD,$\Upsilon=0.72$ more likely than ISO,$\Upsilon=0.52$    \\
\hline
\end{tabular} 
~\\ Comments: ${\rm AIC}=2N+\chi^2$ is the Akaike Information Criterion \citep{Akaike1974}, where $N$ is the number of parameters to fit.
 AIC residuals have been normalized to 100 for clarity. Results are for the standard (STD) interpolation function only, more likely than the simple $\mu-$function.
\end{center}
\end{table*}

\subsection{Results and analysis}  

We  performed non-linear Levenberg-Marquardt least-square fits of the ISO, NFW and MOND models. 
The ISO and NFW fits have two free parameters at fixed $\Upsilon$, three when $\Upsilon$ is left free.
 The MOND fits have one free parameter at fixed $\Upsilon$, and two at free mass-to-light ratio. 
 A normal weighting function set to  $\rm \Delta V_{rot}^{-2}$ is used. 
 Figures~\ref{fig:massmoddm} and~\ref{fig:massmodnodm} show the mass models and Tables~\ref{tab:resmm} list the fitted parameters. 
 A drawback of using different datasets to make an hybrid rotation curve is that 
the resulting distribution of velocity uncertainties is rarely Gaussian. 
We measure for instance a median \ha\ velocity uncertainty that 
is $\sim60\%$ that of the median \hi\ velocity uncertainty.     
A consequence of the error non-homogeneity, combined with the non-uniform sampling of the rotation curve, is to yield 
large $\chi^2$ (reduced values $\gtrsim 6$), which, taken individually, means that a fit has no statistical  significance 
despite of the $\sim 350$ degrees-of-freedom. 
  We thus estimated the differences of $\chi^2$, relatively to a given model, that is, 
 $\Delta \chi^2=(\chi_i^2-\chi_j^2)/\chi_i^2$ 
 to compare the models $i$ and $j$ with $N$ free parameters,  
 and find those that are more likely \citep[see also][]{Martinsson2013}.
 The $\Delta \chi^2$ values are reported in Tabs~\ref{tab:deltachi2}. 

As in many studies of galaxy mass distribution  from fitting of high-resolution rotation curves, 
the M33 mass models were not successful in reproducing all the irregularities of the rotation curve. This is explained by the 
impossibility of our \textit{axisymmetric} modeling and inputs (spherical DM densities, planar gas and surface density profiles) to mimic 
the \textit{asymmetries} in the disks (spiral arms, warp, lopsidedness) that reflect in 
wiggles in the \textit{axisymmetric} rotation curve. 

\subsubsection{Results for dark matter models}
\label{sec:resdm}
The analysis of the residual rotation velocities (bottom insert in Fig.~\ref{fig:massmoddm}) shows that
 NFW halo reproduces more correctly the inner $R=2$ kpc of the rotation curve than 
the core-dominated model, which later  implies 
a total velocity model always smaller than the observation, irrespective of the value of $\Upsilon$. 
An opposite trend is observed  within $R=9-16$ kpc for $\Upsilon=0.52$ 
and $R=9-18$ kpc for $\Upsilon=0.72$, where 
the ISO halo is more appropriate than the cusp, which later implies a total model 
always smaller than the rotation curve. 
At intermediate radii ($2 < R < 6.5$ kpc) and 
in the outer disk ($R\gtrsim 19$ kpc), 
the rotation curve is reproduced equivalently by the cusp and core models.  
None of the models can reproduce the rotation curve for $R=6.5-8$ kpc, where the warp starts. 

In a more global framework, the analysis of $\Delta \chi^2$ first shows that the density shape which is 
 more approriate  at $\Upsilon=0.52$ is that of the NFW halo, 
 while at $\Upsilon=0.72$ it is that of the core-dominated halo. 
 Then, going from $\Upsilon=0.52$ to $\Upsilon=0.72$ improves the modeling for the ISO halo, 
 while it is the opposite way from $\Upsilon=0.72$ to $\Upsilon=0.52$ that  the
 modeling has been improved for the NFW halo.   
 Therefore,  the NFW halo can only coexist with
  $\Upsilon=0.52$ (lower stellar mass) whereas  the ISO model can only coexist with $\Upsilon=0.72$ (higher stellar mass). 
  The models compensate
  by a cuspier density from 0.72 to 0.52 to get a good fit in the central regions.
  This trend is  confirmed by the  free $\Upsilon$ (best-fit) models, 
  as $\Upsilon=0.56$ is found for NFW (stellar mass of $5.9\ 10^9$ \msol) and 
  $\Upsilon=0.8$ is found for ISO (stellar mass of $8.5\ 10^9$ \msol). 
    Note also the  anti-correlation of the cusp concentration with
    the stellar mass (the larger the mass, the smaller the concentration).

 We also deduce from $\Delta \chi^2$ that the combination $\Upsilon=0.52$/NFW  is more likely than the combination
  $\Upsilon=0.72$/ISO.
  The NFW halo as more appropriate model is also apparent at free mass-to-light ratio. 
  Consequently, our mass models are 
  more consistent with 
    the SPS modeling of \citet{Corbelli2014} than with the one used
    by Kam15. A stellar mass-to-light ratio as large as 
    $\Upsilon=0.72$ is therefore excluded for M33.
  
  We therefore adopt the cusp obtained with $\Upsilon=0.52$ as the most likely dark matter halo   for M33, since this mass-to-light ratio has 
  a physical meaning, as based on SPS models described in \citet{Corbelli2014}. The results obtained at free $\Upsilon$ can then be seen as 
  a way to confirm the $\Upsilon=0.52$ results. 
  With  $\Upsilon=0.52$, $\rm V_{200} = 114$ \kms, $c=6.1$, 
  the inferred mass of Messier 33 is $\log (M_{R\le168} /\rm M_\odot) = 11.72$ 
   within the virial radius of the cusp, $R=R_{200}=168$ kpc.
   This total mass estimate is  roughly half those 
  of the Milky Way or  the Andromeda galaxy \citep[][and references therein]{Chemin2009,Bland2016}. 
   The implied baryonic fraction is 2\% at that radius, which strongly differs 
  from the cosmic value, $\Omega_b/\Omega_m = 15.7\%$ \citep{planck16}. This estimate does not include the 
  unknown mass of warm and hot gas. 
   It thus points out that the M33 virial radius cannot be as large as $R_{200}$, unless M33 violates the cosmic value.  
 
    Within a radius of $R=23$ kpc, at the last point of the rotation curve, the total mass is 
   $\log \left(M_{R\le 23}/\rm M_\odot \right) = 10.90$ (or $\sim 13$ times less massive than the Galaxy or M31). The mass fraction of baryons 
   is 11\%, which is more consistent with the cosmic value. 
  This implies that if one assumes that the baryonic fraction of M33 and the cosmic value must be similar, then 
   $R \sim 23$ kpc is close to the real location  enclosing the real total mass of M33. By integrating the density of the adopted NFW halo, we estimate 
   that $R=17-18$ kpc is the radius where the M33 baryonic mass fraction equals the cosmic value.  Interestingly,  this location is  
  very close to the radius where the rotation curve starts to drop ($R \sim 19$ kpc), 
  whose characteristic is expected to occur beyond  the radius that encompasses the total galaxy mass.

   \subsubsection{Results for MOND}     
  All MOND models strongly favour the original, standard interpolation function  over the more 
  simple one (Tab.~\ref{tab:deltachi2}). Figure~\ref{fig:massmodnodm} thus only presents results 
  obtained with the standard interpolation function.
  Moreover, configurations with $\Upsilon=0.52$ are preferred over $\Upsilon=0.72$. 
 This latter model obviously fails at reproducing the rotation curve from $R=9$ kpc.
  This result is reflected in a best-fit mass-to-light ratio of 0.5 for the preferred, standard $\mu$-function, 
  which corresponds to a stellar mass of $5.3\ 10^9$ \msol. 
  The scale acceleration $a_0=1.76\ 10^{-8}$ cm s$^{-2}$ is by 78\%
  larger than the standard value of $a_0= 1.2\ 10^{-8} \times (\rm H_0/75)^2 $ cm s$^{-2}$, for $\rm H_0=68$ \kms\ Mpc$^{-1}$.
    
  The comparison between MOND and the ISO or NFW models can not be done directly 
  from $\chi^2$ residuals  because these models have not the same number of free parameters.  Instead, we made use of 
   Akaike Information  Criterion \citep[AIC,][]{Akaike1974} because 
   it is a simple, linear combination of  both the number of free 
   parameters, $N$, and the fits quality, i.e. ${\rm AIC}=2N+\chi^2$.  Consequently, comparing AIC residuals  
    $\rm \Delta AIC=AIC_{MOND}-AIC_{DM}$ is well appropriate 
    to find which of  MOND and ISO or NFW is more likely than the other 
    \citep[see][]{Chemin2011,Chemin2016}. The
    AIC residuals are reported in Tab.~\ref{tab:monvsdm} for the standard $\mu$-function. 
   The meaning of  $\rm \Delta AIC$ is similar to that of $\Delta \chi^2$. Negative residuals 
    mean MOND more likely than dark matter models, while positive residuals imply MOND less likely than dark 
    matter models. 
    It is found that at fixed mass-to-light ratio $\Upsilon=0.52$, the NFW model is more appropriate than
    MOND, which is more likely than the ISO model. At $\Upsilon=0.72$, the opposite result is found, 
    the ISO model is more appropriate than MOND, which is more likely than the cusp. 
    Another important point is that at free $\Upsilon$, both DM haloes are 
    more appropriate than MOND.  In other words, a more elaborated form than MOND, 
    with a density law of hidden mass and an additional free parameter 
    have had a significant impact on the modeling of the mass distribution of M33. 
    This result is even more significant for NFW than for the pseudo-isothermal sphere.
    This does not imply that MOND has to be rejected, however. This simply illustrates the preference 
    for DM-based models to explain the mass distribution underlying the rotation curve of M33.

 \subsection{Comparison with previous works}
   The finding of a dependency of the inner shape of the DM halo on the mass-to-light ratio is in agreement
    with the analysis of \citet{hague2015}. These authors used a Bayesian approach to constrain the mass distribution of M33 
    from the kinematics of \citet{Corbelli2000}. They excluded a combination of inner DM density slopes shallower 
    than 0.9 with $\Upsilon < 2$.  Their modeling however differs from ours since they used an 
    additional low mass bulge component, and found more likely models having density slopes steeper 
    than the NFW cusp, and with $\Upsilon \sim 1.5$. 
    Such large $\Upsilon$ can only be consistent 
    with shallow DM density profiles with our higher-resolution data, 
    which is ruled out by the present analysis, or by analysis of larger 
    galaxy samples \citep[e.g.][]{Lelli2016}.
    
    At fixed stellar mass, finding a combination  $\Upsilon=0.52$-NFW halo 
    as the most likely result is in very good agreement 
   with the model of \citet{Corbelli2014}. 
   The corresponding halo concentration ($c=6.1$) also agrees with the  value given by these authors ($c=6.7$).
   This concentration is nevertheless not consistent with $c = 9$, which is  the value  inferred from 
   the  Millenium Simulation halo mass-concentration relationship \citep{Ludlow2014} at a redshift of $z=0$ and for our
    virial mass $M_{\rm 200}$. 

   At free stellar mass, the inferred  stellar mass is $5.9\ 10^9$ \msol\  
  (corresponding to a maximum velocity of  73 \kms), which 
    is  23\% larger than the most likely mass of \citet{Corbelli2014} 
    (corresponding to a maximum velocity of 60 \kms). Such a difference is not significant, however, 
    owing to the range of stellar mass predicted by their SPS models.   
     The  halo concentration $c=4.6$ 
    still disagrees with the halo-mass concentration relation, but also with the most likely concentration   given in
   \citet[][$c=9.5\pm 1.5$]{Corbelli2014}.
     The most likely values of \citet{Corbelli2014} were obtained by combining the 
    probability density functions of best-fitting of their rotation curve,  
    a stellar mass compatible with SPS models, and a concentration compatible with the halo mass-concentration relation.  
    
     We verified that the concentration discrepancy  is not caused by  the choice of the adopted higher-resolution
     rotation curve by fitting NFW models at fixed or free stellar mass, and to various rotation curves. 
    Such curves either combined our outer ($R>6.5$ kpc) DRAO velocities with the inner ($R<6.5$ kpc) curve from \citet{Corbelli2014}, or
     our inner \ha\ velocities with the outer GBT points from \citet{Corbelli2014}. 
     All the fits yielded $c \lesssim 7.3$ and a stellar disk mass comparable to our estimate. 
    We also performed a model at fixed concentration $c=9.5$ and free mass-to-light ratio and found 
   $\rm V_{200}=99$ \kms\ and $\Upsilon=0.32$. This corresponds 
    to a maximum velocity of 55 \kms\ for the stellar contribution. 
    It is another way to illustrate the halo concentration-stellar mass degeneracy shown in Sect.~\ref{sec:resdm} and in \citet{hague2015}. 

    To summarize,  the origin of the concentration difference between both  studies  
    can only be the composite likelihood assumption made in \citet{Corbelli2014}. 
    Our higher resolution dataset and adopted cusp model not tied to the halo mass-concentration relation  
    strongly favours a concentration in disagreement with $\Lambda$CDM simulations.  
    Since the stellar mass is the key parameter that allows the concentration to match the NFW cusp with the central mass 
    distribution imposed by the rotation curve, it is only once it is 
    known with more accuracy that the cusp concentration conflict will be alleviated in M33. 

%% file: Conclusion.tex
New high sensitivity \hi\ observations of M33 obtained with the DRAO interferometer  have been presented. Combined with
the single dish Arecibo data from \citet{Putman2009}, the dataset reach column densities 
as low as $\sim 5~ \times ~10^{18}$ cm$^{-2}$ in the outer \hi\ disk of M33. 

The main results on the \hi\ distribution and kinematics of M33 are:
 
 \begin{itemize}
 
\item While the bulk of the \hi\ gas is found within the stellar disk ($\le
R_{25} \sim 8$ kpc), the    \hi\ distribution is traced out to $\ge  2.7 R_{25}$. It is irregular in the outer region, in the form of tails or arc-like features or isolated clumps. 
More gas is detected in the Southern receding disk side, and the \hi\ emission is more extended 
to the North-West than to the South-East of the disk, implying a lopsided \hi\ distribution. 
The surface density is nearly constant out to the edge of the stellar disk, and then drops abruptly. 
At the adopted distance of 840~kpc, the \hi\ mass is $\sim 2 \times 10^9$~\msol\  for a  $\rm M_{HI}/ L_{V} \simeq 0.2$. 

\item Position-velocity diagrams make it possible to evidence ``beard-like'' contours from a low brightness component with  
an important velocity scatter, as it is observed  lagging and exceeding the disk rotation, as well as leaking in the 
forbidden velocity zone of apparent counter-rotation.  

\item The rotation curve is in very good agreement with the \ha\ curve of \citet{ZKam2015}
 in the inner disk, and consistent with the \hi\ rotation curve of \citet{Corbelli2014} inside $R=20$ kpc.
Beyond that radius,  the \hi\ rotation curves of the approaching and receding disk sides differ by up to 60 \kms.

\item The warp of M33 consists mainly in a strong twist 
of the position angle of the kinematical major axis ($\sim 40\degr$) beyond $R = 7$ kpc. 
This result is in perfect consistency with previous \hi\ studies.
Only a minor increase of inclination is detected throughout the entire
 disk ($\sim 5\degr$). 

\item Wider and double-peaked \hi\ profiles are evidenced  in a large-scale, incomplete ring-like structure of larger dispersion.
They coincide with the transition zone of twist of the major axis position angle between the inner and outer regions. 
The crowding of rings inferred by our warp model naturally explains part of the larger dispersion and multiple peaks. 
Collisions of gas clouds are expected in this region, and the gas orbits are likely elongated in the direction 
to the companion Messier 31. Other wider \hi\ profiles that are not in the crowded rings zone are associated to 
holes in the \hi\ distribution.

\item A Fourier series analysis of the velocity field reveal non-circular and asymmetric motions, suggesting perturbations of the first and second order of the gravitational 
potential of M33. The asymmetric motions are all observed to increase
in the disk outskirts.  

\item The past tidal interaction with Messier 31 already  evidenced by  large-scale gas and stellar surveys 
of the M33 enviromnent \citep[e.g.][]{Braun2004,Putman2009,Ibata2014}
is likely at the origin of most of the perturbed \hi\ kinematics and morphology presented here. 

\end{itemize}

The main results of the M33 mass distribution modeling from the hybrid \ha-\hi\ rotation curve  are:

\begin{itemize}
 
\item The most likely density shape of dark matter is cuspy, to the detriment of a pseudo-isothermal sphere. 
The concentration of the most likely NFW halo disagrees with that expected by the halo mass-concentration from CDM numerical models, 
or from previous \hi\ studies.  
Modified Newtonian Dynamics is less likely than models with a dark matter halo.
 
\item The mass enclosed within the virial radius of the best-fit NFW halo (168 kpc) is $\sim 5.2\, 10^{11}$ \msol, 
implying a very low baryonic mass fraction (2\%), in conflict with the  
cosmic value of 15.7\%. That result suggests a more plausible M33 virial radius well smaller than that of the adopted cusp. 
 
  \item  The most likely mass of the stellar disk is $5.5\, 10^{9}$  \msol, only about 3 times larger 
 than the mass of disk of neutral Hydrogen.
 The enclosed mass within $R = 23$ kpc  
 at the last point of the rotation curve is $\sim 7.9\, 10^{10}$  \msol.
  Luminous matter represents about 11\% of that mass, in better agreement with the cosmic value. A radius as 
low as the radius of the \hi\ disk could thus be very nearby the true  location encompassing the total mass of M33. 

\end{itemize}